\documentclass{aa}

\usepackage{graphicx}
\usepackage{txfonts}
\usepackage{caption}
\usepackage{subcaption}
\usepackage{tablefootnote}
\usepackage{hyperref}
\usepackage{comment}
\usepackage{xcolor}
\usepackage[normalem]{ulem}

\newcommand{\mpo}[1]{\textcolor{red}{[MP:OK]}}

\begin{document}

   \title{Magnetically Driven Obliquity in Circumplanetary Disks and Twisted Bipolar-jet Formation}
   \titlerunning{Magnetically Driven Obliquity in Circumplanetary Disks}

   \author{Raúl O. Chametla 
          \inst{1}
          \and 
          Martin E. Pessah\inst{2}
          \and 
          F. J. S\'anchez-Salcedo\inst{3}
          \and
          David Vokrouhlick\'{y}\inst{1}
          \and
          Mauricio Reyes-Ruiz\inst{4}
          \and
          Ond\v{r}ej Chrenko\inst{1}
          \and
          Alessandro Morbidelli\inst{5,6}
          }
   \authorrunning{R.~O.~Chametla et al.}
   \institute{%
            Charles University, Faculty of Mathematics and Physics, Astronomical Institute,
            V Hole\v{s}ovi\v{c}k\'ach 747/2, 180 00,
            Prague 8, Czech Republic\\
            \email{raul@sirrah.troja.mff.cuni.cz}
             \label{UKarlova}
         \and
            Niels Bohr International Academy, Niels Bohr Institute, Blegdamsvej 17, DK-2100 Copenhagen Ø, Denmark
             \label{NBIA}
         \and
            Instituto de Astronomía, Universidad Nacional Autónoma de México, 
            Apt.~Postal 70-264, C.P.~04510, Mexico City, Mexico
             \label{UNAM}
        \and
            Instituto de Astronomía, Universidad Nacional Autónoma de México, Ensenada, 22800 B.C., México\label{UNAMIAE}
        \and
            Université Côte d’Azur, Observatoire de la Côte d’Azur, CNRS, Laboratoire Lagrange, Nice, France\label{NICE}
        \and
            Collège de France, CNRS, PSL Univ., Sorbonne Univ., Paris, 75014, France\label{CNRS}
             }

   \date{Received XXX; accepted YYY}

\abstract{Circumplanetary disks (CPDs) regulate gas accretion onto forming giant planets and provide the environment in which their satellites may form. We use high-resolution, global three-dimensional simulations to investigate the early formation, orientation, and outflows of a CPD around a Jupiter-mass planet embedded in a turbulent magnetized protoplanetary disk. Within a locally isothermal, ideal-MHD framework, we evolve disks threaded by net vertical magnetic fields, corresponding to initial plasma parameters $875\leq\beta\leq3500$, until magnetorotational-instability-driven turbulence is established before inserting the planet. We also perform a hydrodynamic control simulation. In the most strongly magnetized model, with $\beta=875$, the CPD forms already highly inclined and reaches a maximum tilt of approximately $87^\circ$. By contrast, the hydrodynamic CPD and the MHD models with $\beta\gtrsim1000$ remain nearly coplanar. A control simulation in which the planet is inserted before global MRI turbulence develops also remains coplanar, despite producing local turbulence and bipolar outflows. The large tilt is therefore associated with the pre-existing global turbulent state and its evolved velocity and toroidal magnetic-field structure, although our current diagnostics do not distinguish between a direct magnetic torque and the accretion of misaligned angular momentum. All MHD models launch bipolar outflows; in the highly tilted case, these develop a curved, helical morphology that persists until the end of our short-term simulations. These results identify pre-existing global magnetized turbulence as a viable route to generating strongly inclined CPDs and twisted planetary outflows.}


   \maketitle
%

\section{Introduction}
\label{sec:introduction}

In the core-accretion scenario for the formation of a Jupiter-mass planet \citep{Pollack1996}, the quasi-static contraction and runaway gas-accretion stages may be accompanied by gas accretion through a circumplanetary disk (CPD) that forms within the planet's Hill sphere. CPDs are of interest in their own right because they provide natural sites for satellite formation around giant planets \citep[see for instance,][and references therein]{BatM2020ApJ}. However, understanding the formation and evolution of CPDs remains challenging because their geometry and boundary conditions are set jointly by the properties of the planet and their interaction with the surrounding protoplanetary disk, while gas thermodynamics, stellar irradiation, and magnetic fields also shape their dynamics. Several studies have focused on CPD obliquity \citep[e.g.,][]{Mar2020,Mar2021,MarA2021} to understand the coupled evolution of the CPD and planetary spin orientations.%
\footnote{Alongside studies of the early evolution of obliquity in CPD--planet systems, a broad literature examines dynamical pathways for tilting planetary spins that generally operate well after the planet-formation phase.}
These models generally require a process capable of generating or sustaining the misalignment, such as stochastic angular-momentum delivery through turbulent gas accretion \citep[e.g.,][]{Mar2020}. Here, we investigate whether a CPD can instead form misaligned when a planet is embedded in an MRI-turbulent, magnetized protoplanetary disk.

Over the past decades, the complexity of this problem has motivated a progression from highly simplified treatments focusing on individual aspects to increasingly comprehensive models, enabled in part by advances in computational power. Early one-dimensional models \citep{Mizuno1980} were followed by two-dimensional simulations \citep{Kley1999,Lubow1999,G2002} and, subsequently, by three-dimensional studies. These models incorporate a range of physical ingredients, from purely hydrodynamic treatments \citep{G2003,Tanigawa2012} to approaches that include radiative transfer \citep{Szulagyi2016,Marleau2023,Krapp2024,Lega2024}.

\begin{figure*}
    \centering
    \includegraphics[width=1.0\linewidth]{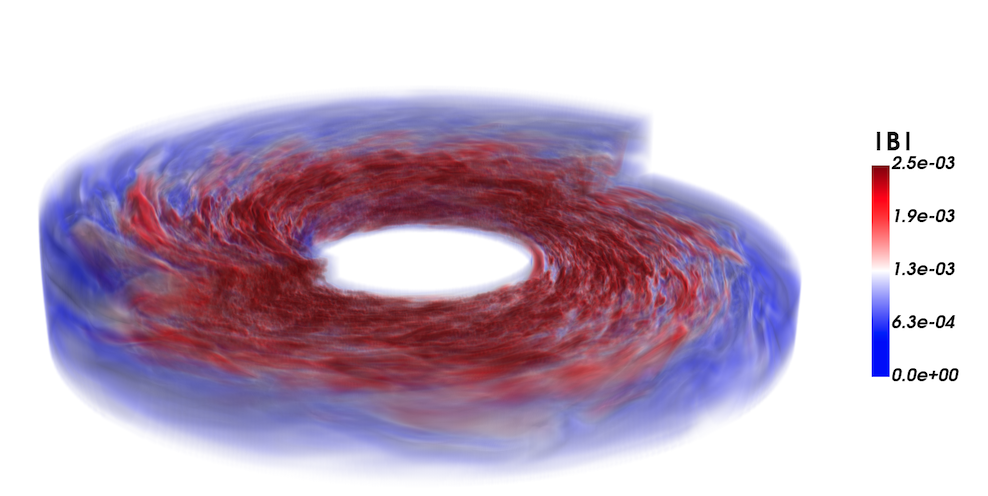}
    \caption{Three-dimensional view of the magnetic-field magnitude $|\mathbf{B}|$ in the protoplanetary disk at $t'=8$ orbits. The disk thickness has been stretched by a factor of 2 for visualization. The planet is subsequently inserted into this turbulent disk state, which serves as the initial condition for CPD formation.}
    \label{fig:disk}
\end{figure*}

\begin{figure}
 \centering
 \includegraphics[width=0.9\linewidth]{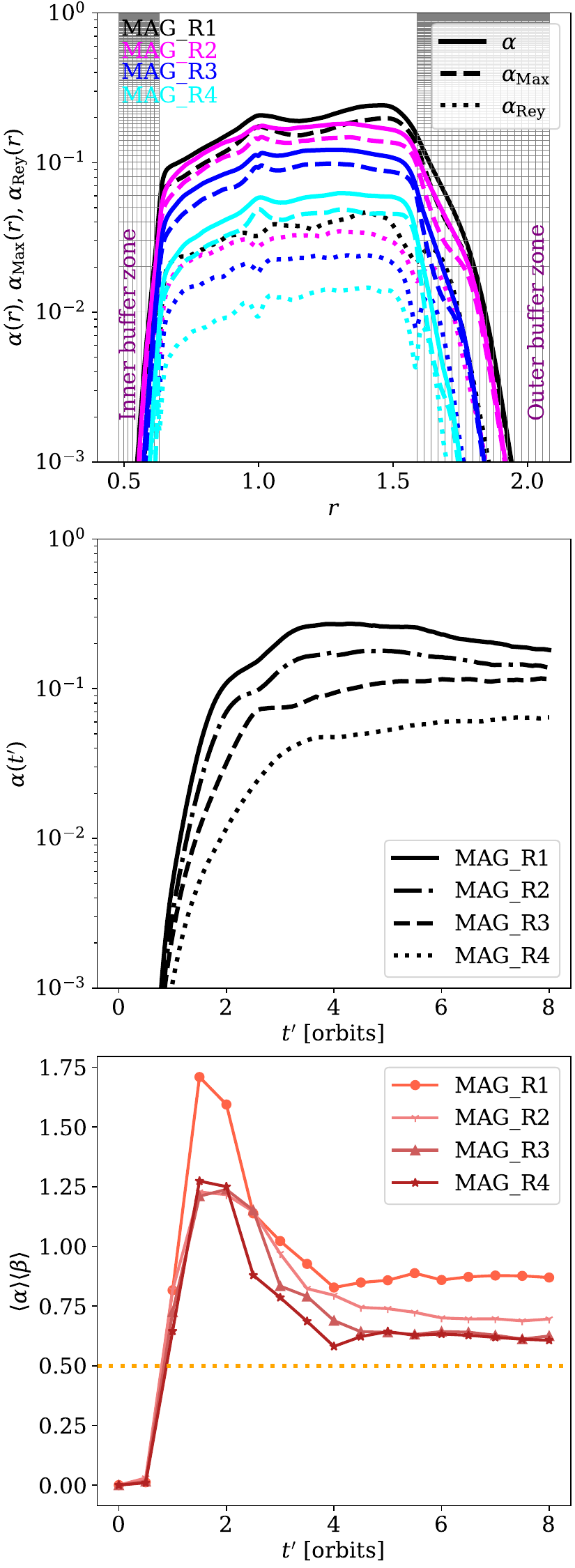}
\caption{Turbulence parameters in the disk before planet insertion. Top panel: Radial profiles of the total $\alpha$ parameter and its Maxwell ($\alpha _{\mathrm{Max}}$) and Reynolds ($\alpha _{\mathrm{Rey}}$) contributions, defined in Eqs.~\ref{eq:alpha}--\ref{eq:alphaM}, at $t'=8T_0$. Middle panel: Temporal evolution of the radially averaged $\alpha$ parameter. Bottom panel: Temporal evolution of the product $\langle\alpha\rangle\langle\beta\rangle$, where the angle brackets denote volume averages.}
 \label{fig:ic}
\end{figure} 

Gas accretion onto the planet is regulated by the distribution and transport of angular momentum. Among the relevant processes, the magnetorotational instability \citep[MRI;][]{BH1991,BH1998}, which can drive turbulence in differentially rotating, magnetized flows, plays an important role. It is therefore important to study how magnetic fields influence the structure and geometry of CPDs. For instance, recent studies incorporating non-ideal magnetohydrodynamics have shown that weak coupling between the gas and magnetic field may significantly reduce the efficiency of angular momentum transport in CPDs \citep{LM2013,Fujii2017}. Magnetic fields may also enable the launching of outflows from the vicinity of forming planets \citep{Gressel2013,GW2023}.

Global three-dimensional magnetohydrodynamic (MHD) simulations that include non-ideal effects provide a more realistic description of the planet-induced gap in the protoplanetary disk, as well as of CPD formation and evolution. However, their computational cost can make systematic explorations of long-term evolution prohibitively expensive \citep[see][]{Gressel2013}. Here, we address this difficulty by considering CPD formation around a planet embedded in a turbulent, locally isothermal protoplanetary disk. The locally isothermal treatment favors the formation of a CPD rather than an envelope \citep[see][]{Krapp2024}, while adopting ideal MHD allows us to study its structure and evolution at a reduced computational cost and higher spatial resolution. Our main objectives are (i) to analyze CPD formation in a turbulent environment and determine its resulting structure, geometry, and orientation, and (ii) to identify the conditions under which twisted bipolar planetary jets form.

The paper is organized as follows. In Section~\ref{sec:equations}, we present the mathematical setup and numerical implementation of our global 3D MHD simulations. In Section~\ref{sec:results}, we discuss the main results in the strong-field regime. Section~\ref{sec:discussion} examines the physical mechanisms underlying these results and briefly explores their dependence on magnetic field strength, parametrized by the plasma parameter $\beta$. Finally, Section~\ref{sec:conclusions} summarizes our conclusions and the broader implications of this work.

\begin{table}
\caption{Overview of numerical models.}         
\label{table:1}     
\centering                    
\begin{tabular}{c c c}     
\hline\hline                      
Name & Magnetic field $B_z$ (mG) & Plasma parameter $\beta$ \\    
\hline                            
    HYDRO & 0.0 & $\infty$ \\    
    MAG\_R1 & 54.0 & 875 \\
    MAG\_R2 & 50.0 & 1000 \\
    MAG\_R3 & 38.16 & 1750 \\
    MAG\_R4 & 27.0 & 3500 \\
\hline                            
\end{tabular}
\tablefoot{The second column lists the initial value of vertical component $B_z$ of the magnetic field in the simulation. The third column provides the plasma parameter $\beta$.}
\label{tab:t1}
\end{table}

\begin{figure*}
 \centering
 \begin{subfigure}{0.45\textwidth}
   \includegraphics[scale=.22]{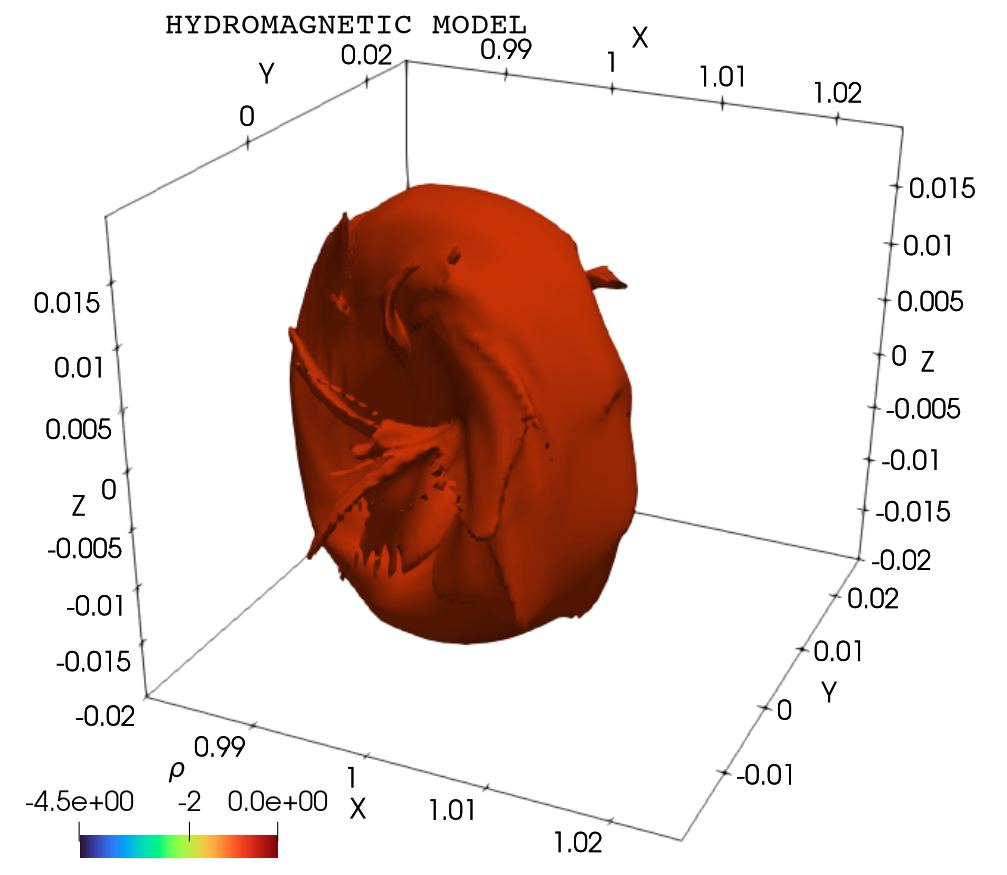}
  \end{subfigure}
  \hfil
  \begin{subfigure}{0.45\textwidth}
    \includegraphics[scale=0.22]{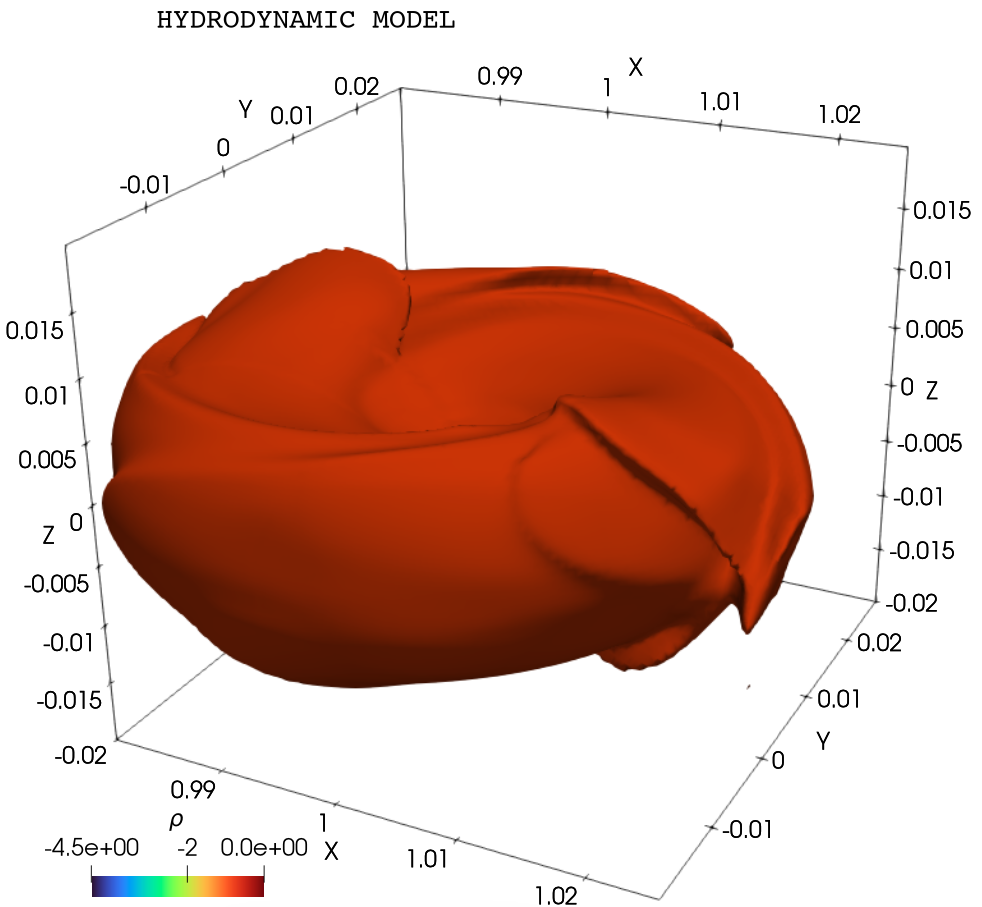}
  \end{subfigure}
\caption{
Three-dimensional isodensity surfaces of the CPDs at $t=5\,T_0$ in the MAG\_R1 model (left panel) and the HYDRO model (right panel; see Table~\ref{tab:t1}). Both panels use the same density threshold. The CPD is highly inclined relative to the protoplanetary disk midplane ($z=0$) in the MAG\_R1 model, whereas it remains aligned with the midplane in the HYDRO model. The isodensity surface also has a larger radial extent in the HYDRO model. In the coordinates shown, the Hill radius is $r_{\rm H}\simeq0.06\,r_p$, with the larger HYDRO CPD extending to approximately one-third of $r_{\rm H}$.} 
\label{fig:comparison_HM}
\end{figure*}

\begin{figure*}
 \centering
 \includegraphics[width=0.8\linewidth]{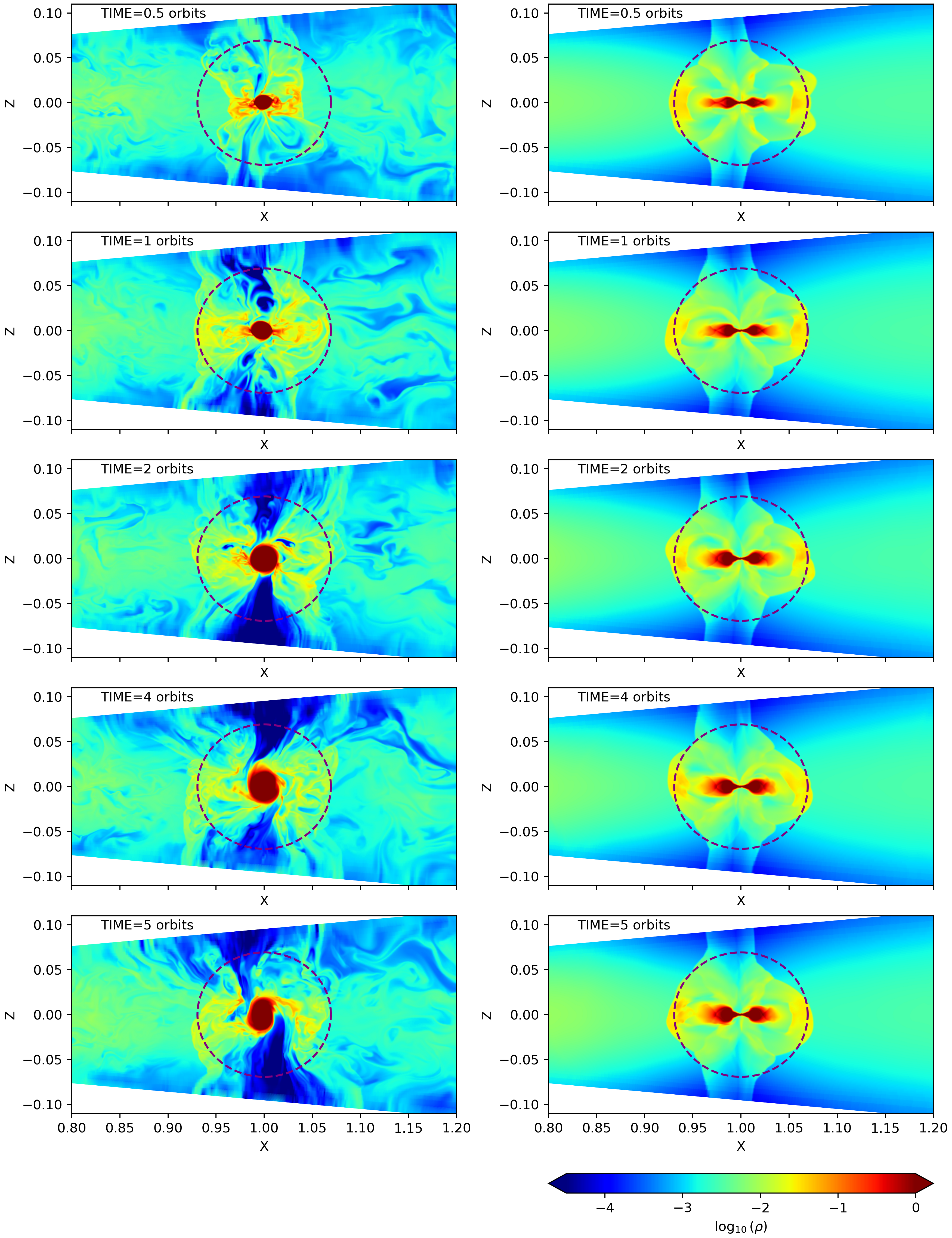}
\caption{Temporal evolution of the gas density around a Jupiter-mass planet in the meridional $X$-$Z$ plane. Left panels: Magnetized CPD in the MAG\_R1 model. Bipolar jets are launched from the vicinity of the planet and extend across the Hill-sphere boundary, denoted by the dashed circle. The central high-density structure corresponds to the intersection of the meridional slice with the tilted CPD. Right panels: Nonmagnetized CPD in the HYDRO model. A well-defined coplanar CPD and gas inflow toward the planet are already apparent by the first orbital period. Outside the Hill sphere, the density distribution remains comparatively smooth, in contrast to the turbulent structure of the magnetized model.}
 \label{fig:jets}
\end{figure*}

\section{Theory}
\label{sec:equations}
Our simulations describe 3D gas flow by employing ideal MHD equations, which in the inertial frame read:
\begin{equation}
\frac{\partial\rho}{\partial t}+\nabla\cdot(\rho \mathbf{v})=0,
\label{eq:gas_cont}
\end{equation}
\begin{equation}
\frac{\partial(\rho\mathbf{v})}{\partial t}+\nabla\cdot(\rho\mathbf{v}\mathbf{v}+p\mathbf{I})= -\rho\nabla\Phi+\nabla\cdot\tau+\mathbf{J}\times\mathbf{B},
\label{eq:gas_mom}
\end{equation}
\begin{equation}
\frac{\partial \mathbf{B}}{\partial t}=\nabla\times(\mathbf{v}\times\mathbf{B}),
    \label{eq:induc}
\end{equation}
where $\rho$ and $\mathbf{v}$ denote the gas density and the gas velocity, respectively, $\Phi$ is the gravitational potential, $\mathbf{I}$ is the unit tensor, and $p$ the gas pressure. In Equations~(\ref{eq:gas_mom}) and~(\ref{eq:induc}), $\mathbf{J}$ and $\mathbf{B}$ are the current density and the magnetic field, respectively. In the non-relativistic limit their relation $\mathbf{J}=(\nabla\times\mathbf{B})/\mu_0$ expresses Amp\`{e}re's law. The magnetic field obeys the solenoidal condition $\nabla\cdot\mathbf{B}=0$. For the gas pressure, we consider the equation of state
\begin{equation}
 p=c_s^2\,\rho, \label{eq:pressure}
\end{equation}
where $c_s$ is the isothermal sound speed. The second term in the right-hand-side of Eq.~(\ref{eq:gas_mom}) is the viscous stress tensor given as
\begin{equation}
\tau=\rho \nu\left[\nabla\mathbf{v}+(\nabla\mathbf{v})^{T}-\frac{2}{3}(\nabla\cdot\mathbf{v})\,\mathbf{I}\right],  \label{eq:stresstensor}
\end{equation}
where $\nu=\alpha c_sH$ is the gas dynamic viscosity parameter \citep[e.g.,][]{SS1973}. 
This term is used only in the purely hydrodynamic simulation (in which we set $\mathbf{B}=0$; see Table~\ref{tab:t1}). In the MHD runs, angular momentum is instead transported by turbulent Reynolds and Maxwell stresses, and no explicit viscosity is included. For the HYDRO simulation, we adopt a constant Shakura--Sunyaev viscosity parameter of $\alpha = 10^{-4}$, representing a weakly viscous disk. This choice is motivated by three-dimensional, unstratified, non-ideal MHD shearing-box simulations in the strong-coupling limit, which show that ambipolar diffusion can strongly suppress MRI-driven stresses depending on the coupling strength and magnetic-field geometry \citep{BS2011}. It also lies at the lower end of the effective $\alpha$ values inferred from protoplanetary-disk masses and stellar accretion rates \citep[e.g.,][]{Rav2017}, although such estimates do not necessarily measure local turbulent transport.

\subsection{Magnetic field in the protoplanetary disk}

Using dust polarization as a tracer of magnetic-field geometry in protoplanetary disks, \citet[][]{Ohashi2025} reported the first observationally derived estimates of the relative strengths of the three magnetic-field components in the HD~142527 protoplanetary disk. They inferred a total magnetic-field strength of $\sim 0.3$~mG at a radius of $\sim 200$~au. Motivated by this evidence that dynamically relevant magnetic fields are present in protoplanetary disks, here we adopt an initially uniform vertical magnetic field with a strength between $B_z=27\,\mathrm{mG}$ and $B_z=54\,\mathrm{mG}$ \citep[see Table~\ref{tab:t1}; see also][]{Gressel2013}. These field strengths correspond to a plasma parameter $\beta$ of
\begin{equation}
 \beta\equiv 2\mu_0c_s^2\rho/B^2\in[875,3500]  \label{eq:pla}
\end{equation}
at the reference radius $r_0=1$ (corresponding to $r_0=5.2$~au in physical units). We measure time in units of $T_{0}$, the orbital period at $r_{0}$. Because the planet is introduced at $r_{p}=r_{0}$, we use $r_p$ instead of $r_0$ hereafter.

\subsection{Protoplanetary gas disk set-up}
The setup of our three-dimensional numerical model is based on the gaseous disk model presented by \citet[][see their Appendix A]{MB2016}. We use spherical coordinates $(r,\theta,\phi)$, where $r$ is the radial distance from the star, $\theta$ is the polar angle ($\theta=\pi/2$ at the midplane of the protoplanetary disk) and $\phi$ is the azimuthal angle. We use the reference frame co-rotating with the planet (located at $\phi=0$) and the origin fixed on the central star\footnote{As in \citet{Chametla2026}, we present some results in a planetocentric coordinate system. To distinguish the magnetic-field components defined in this frame from those defined in the star-centered frame, we append the subscript $c$ to the latter.},
which requires additional forces (the centrifugal and Coriolis) on the right-hand-side of Eq.~(\ref{eq:gas_mom}).

The aspect ratio of the gas disk is $h = H/r$, and its volume density is given by
\begin{equation}
 \rho\left(r,\theta\right)=\rho_\mathrm{eq}\left(r\right)\,(\sin\theta)^{-\sigma-\xi+h^{-2}},\label{eq:rhog}
\end{equation}
with 
\begin{equation}
 \rho_\mathrm{eq}\left(r\right)=\frac{\Sigma_0}{\sqrt{2\pi}h r_p}\left(\frac{r}{r_p}\right)^{-\xi},   \label{eq:rhoeq}
\end{equation}
where $\Sigma_0=2\times10^{-3}/\pi$ is the surface density in code units (which corresponds to $\approx200$ g~cm$^{-2}$ at $r_p=5.2$~au). For the models considered here, $\sigma=1$, $\xi=1.5$ and $h=0.05$. We initialize the gas velocity components with $v_r=v_\theta=0$ and 
\begin{equation}
  v_\phi=\sqrt{\frac{GM_\star}{r\sin{\theta}}-\xi c_s^2}. \label{eq:vphi}
\end{equation}
Here, $M_\star$ is the mass of the star taken to be one solar mass.

\subsection{Gravitational potential}
The gravitational potential $\Phi$ is given by
\begin{equation}
\Phi=\Phi_S+\Phi_p,
 \label{eq:potential}
\end{equation}
where
\begin{equation}
\Phi_S=-\frac{GM_\star}{r},
 \label{eq:Star_potential}
\end{equation}
and
\begin{equation}
\Phi_p=-\frac{GM_p}{\sqrt{r'^2+r_\mathrm{sm}^2}}+\frac{GM_p r\cos\phi\sin\theta}{r_p^2}
 \label{eq:Planet_potential}
\end{equation}
are the stellar and planetary contributions, respectively. In Eq.~(\ref{eq:Planet_potential}), $r'=|\mathbf{r}-\mathbf{r}_p|$ is the distance between the grid cell and the planet, and $r_\mathrm{sm}$ is a softening length used to regularize the planetary potential near the planet. The second term on the right-hand side of Eq.~(\ref{eq:Planet_potential}) is the indirect potential term, which accounts for the acceleration of the star-centered reference frame due to the planet. The simulations were performed with $r_\mathrm{sm} = 0.0072~r_{\rm H}$, where $r_{\rm H}=r_p\,(M_p/3M_\star)^{1/3}$ is the Hill radius. This value is comparable to the local grid-cell size.

\subsection{Mesh resolution}

The numerical domain extends from $0.48\,r_{p}$ to $2.08\,r_{p}$ in the radial direction, from $-\pi$ to $\pi$ in the azimuthal direction, and from $\pi/2-2\,h$ to $\pi/2+2\,h$ in colatitude\footnote{Although the high computational cost of generating square cells throughout the planetary Hill sphere required a limited colatitude domain, it remains sufficient to accurately capture both the formation and subsequent twisting of jets near the planet.}. To achieve high resolution inside the Hill sphere, where the CPD forms, we apply the mesh-density-function approach \citep[see][for details]{Bll2023}. In the radial and azimuthal directions, we use the coefficients $a_r=a_\phi=0.2618$, $b_r=b_\phi=0.3141$, $c_r=15.0$, and $c_\phi=1.5$. For colatitude, we use the prescription described in the Appendix of \citet{Chametla2025}, which yields approximately cubic grid cells in the refined region. The mesh refinement is centered on the planet and extends beyond the Hill sphere. Within this region, we maintain a constant grid spacing of $5\times10^{-4}\,r_p$ in all three directions. The full computational domain contains $(N_r,N_\theta,N_\phi)=(1864,256,1864)$ grid zones.

\begin{figure}
 \centering
 \includegraphics[width=1.0\linewidth]{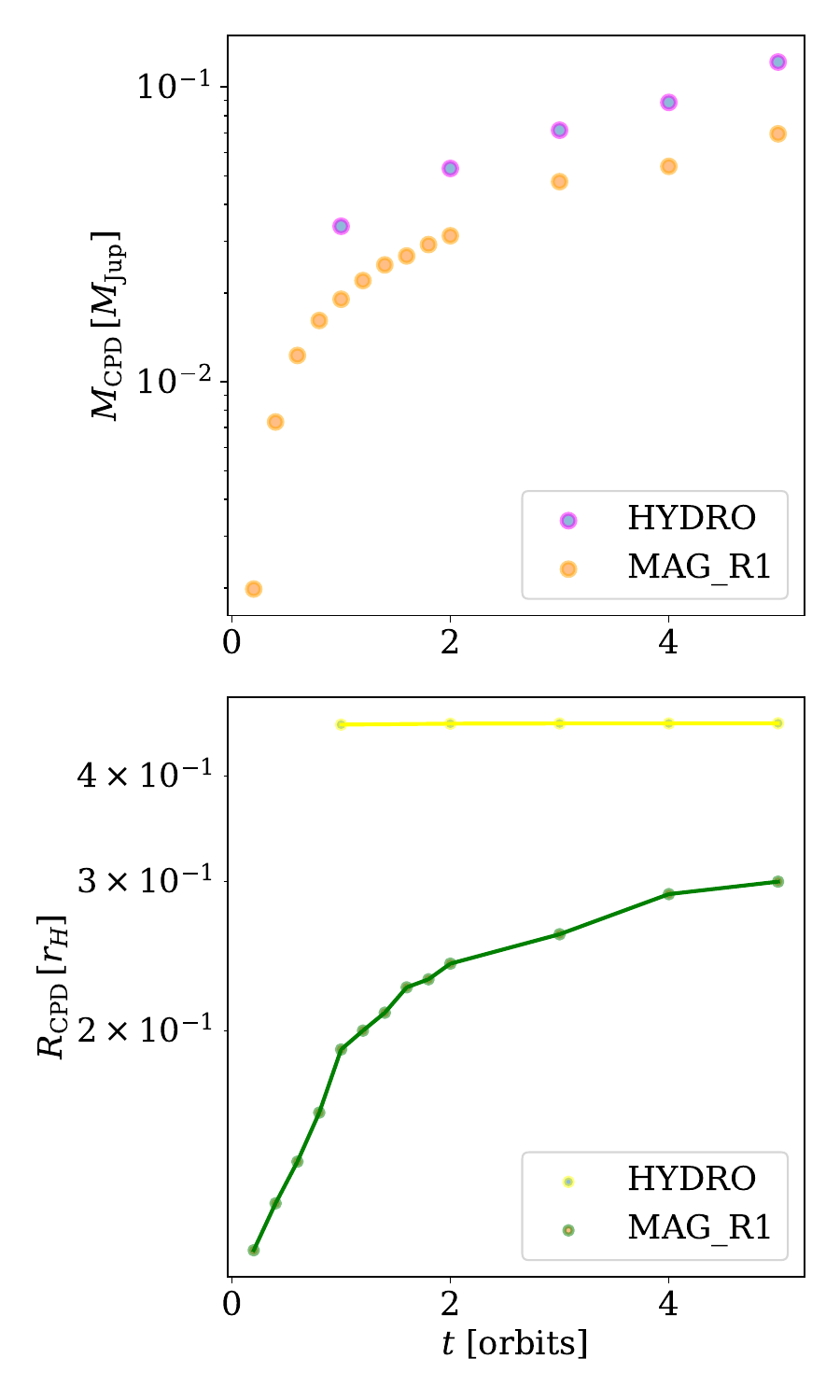}
  \caption{Temporal evolution of the mass (top) and radius (bottom) of the CPD for the HYDRO and MAG\_R1 models.}
 \label{fig:mt}
\end{figure}

\begin{figure}
    \centering
    \includegraphics[width=1.0\linewidth]{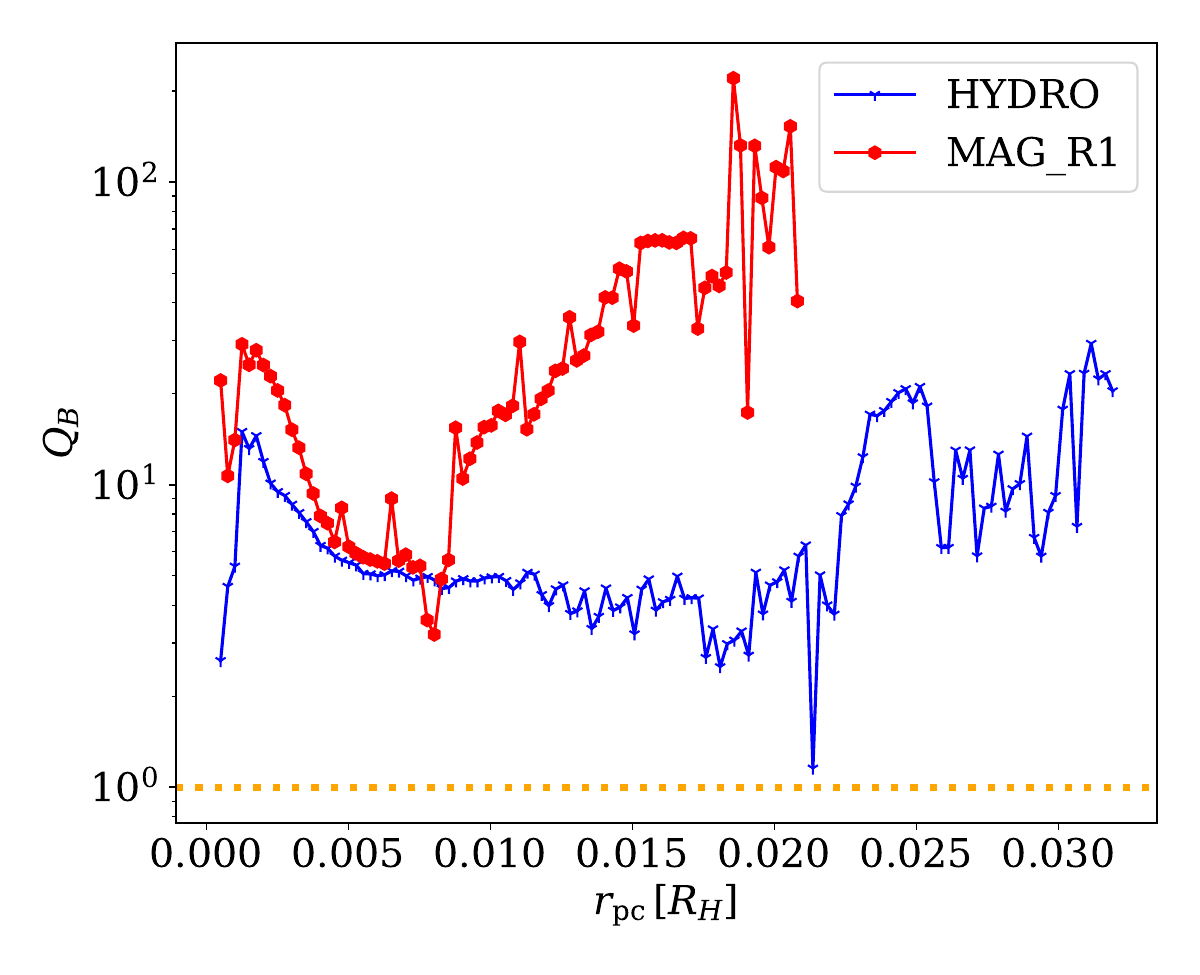}
    \caption{Radial profile of Toomre parameter in the CPD for the HYDRO and MAG$\_$R1 models. Note that in the HYDRO model we set $v_A=0$.}
    \label{fig:Toomre}
\end{figure}

\begin{figure}
 \centering
 \includegraphics[width=0.99\linewidth]{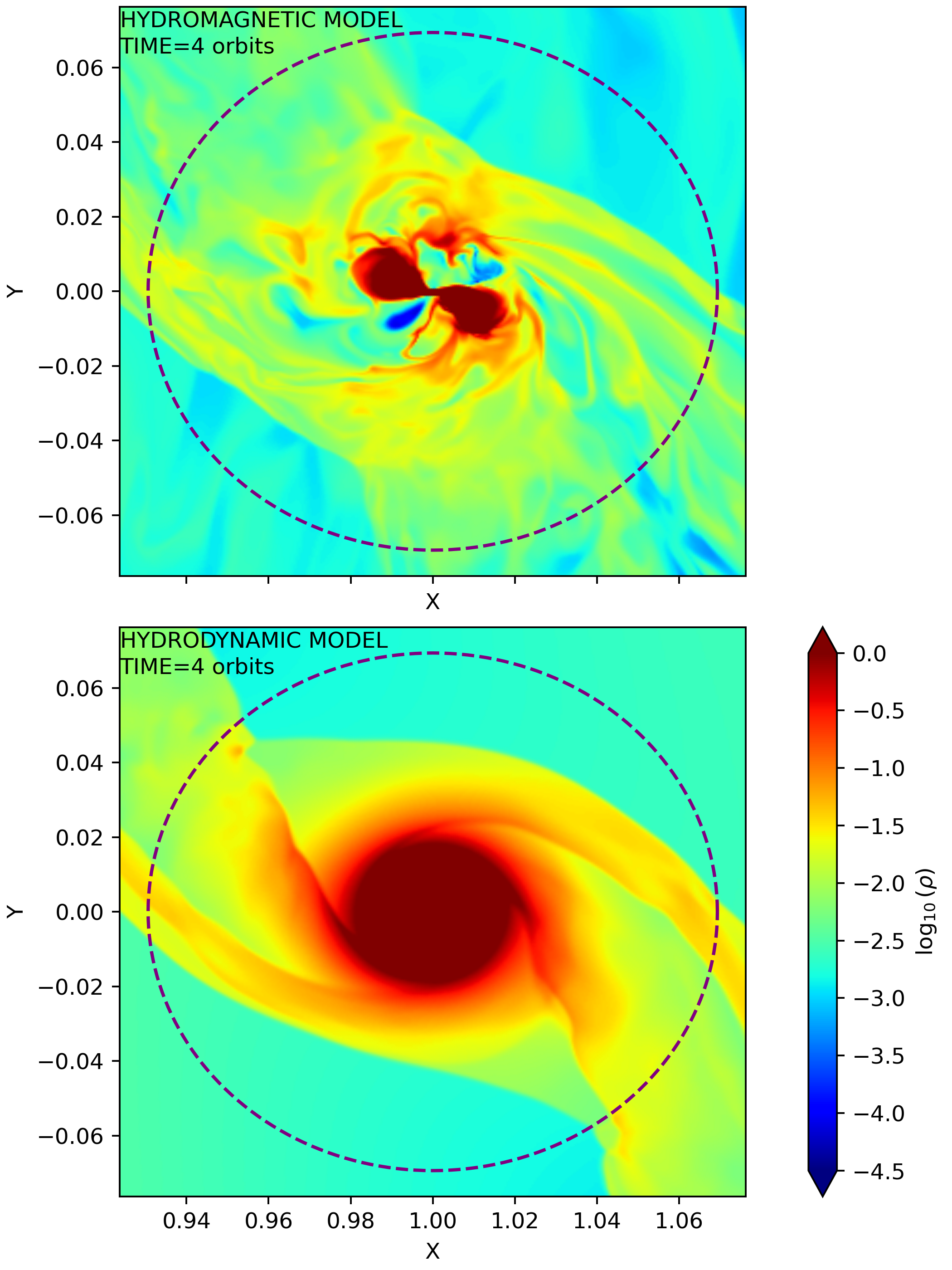}
\caption{Gas density in the protoplanetary disk midplane ($Z=0$) within the planet's Hill radius, whose boundary is marked by the dashed circle. Top panel: MAG\_R1 model. The central region, $r_{pc}<0.5\,r_\mathrm{H}$, contains two pairs of distinct features: (i) two low-density lobes (dark blue), which coincide with the bases of the bipolar jets, and (ii) two high-density lobes (dark red), which arise where the highly tilted CPD intersects the protoplanetary disk midplane. Bottom panel: HYDRO model. The inner region, $r_{pc}<0.5\,r_\mathrm{H}$, displays the characteristic face-on structure of a coplanar CPD, whose rotation axis is aligned with the vertical $Z$-axis.}
    \label{fig:midplane}
\end{figure}

\begin{figure}
 \centering
 \includegraphics[width=1.0\linewidth]{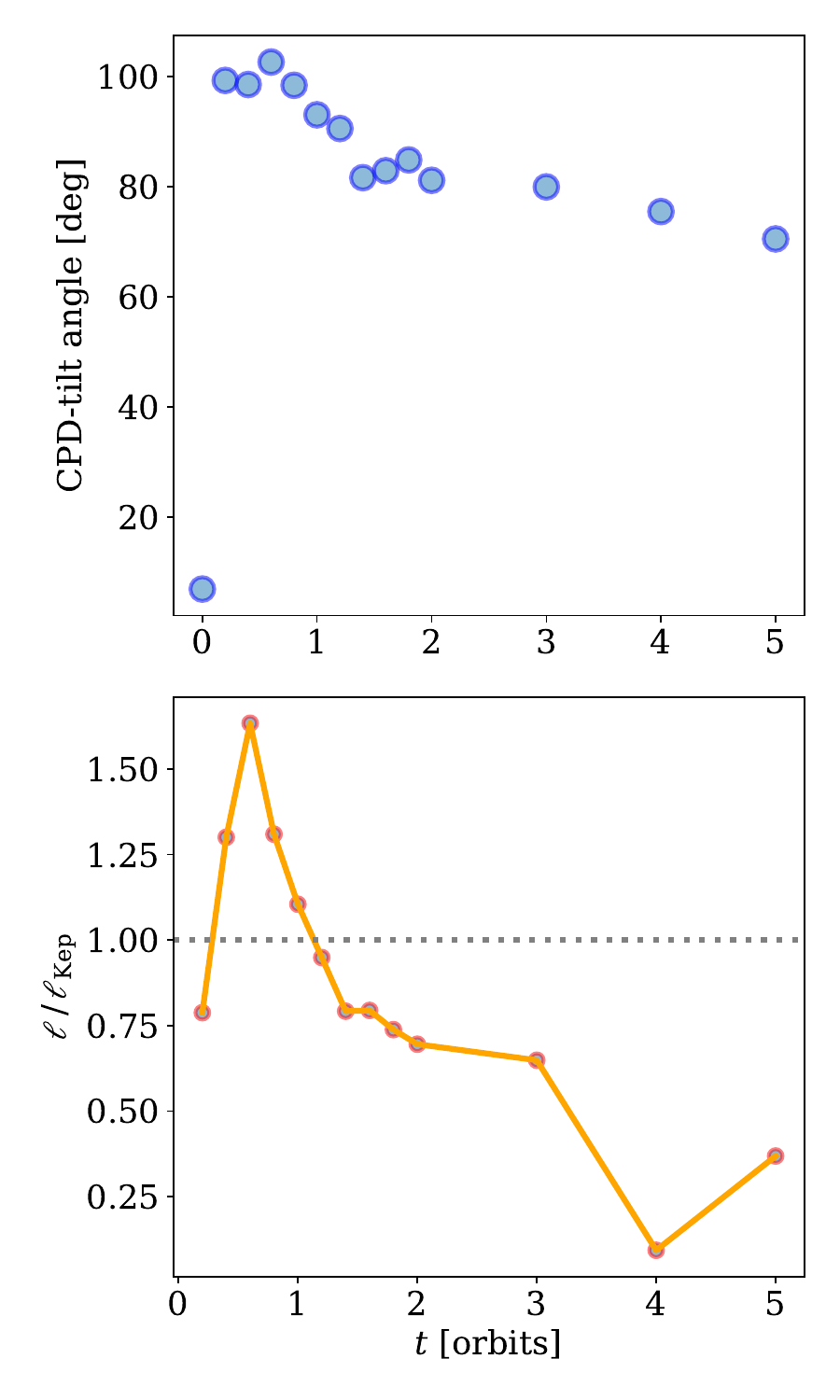}
  \caption{Temporal evolution of the tilt angle of the angular momentum vector of the gas within the Hill sphere (in degrees; top panel) and of the 
  specific angular momentum magnitude (bottom panel). The rapid evolution during the first revolution of the planet is followed by a slower transition towards a quiescent phase.
  }
 \label{fig:tangle}
\end{figure}

\begin{figure*}
 \centering
 \includegraphics[width=1.0\linewidth]{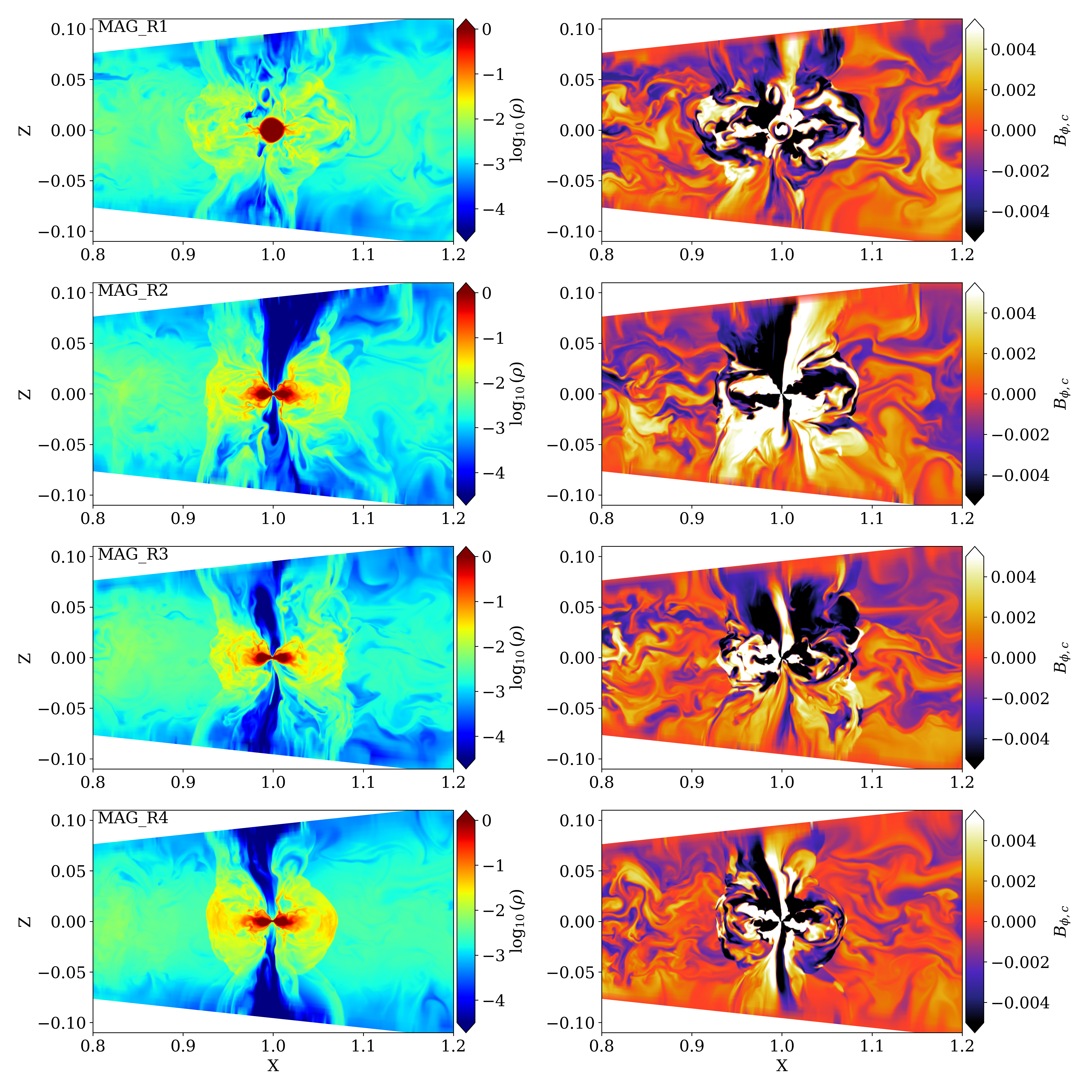}
  \caption{Gas density (left) and azimuthal magnetic field component defined with respect to the central
  star (right) in the vicinity of the planet, in the meridional $X-Z$ plane, at $t=1.5$ orbits, for models MAG\_R1, MAG\_R2, MAG\_R3 and MAG\_R4 (see Table~\ref{tab:t1}). In the latter three models, the weaker magnetic fields produce turbulence that does not lead to a CPD tilt.
   Nevertheless, bipolar jets are formed in all our simulations.}
 \label{fig:new_models}
\end{figure*}

\begin{figure*}
 \centering
 \includegraphics[width=1.0\linewidth]{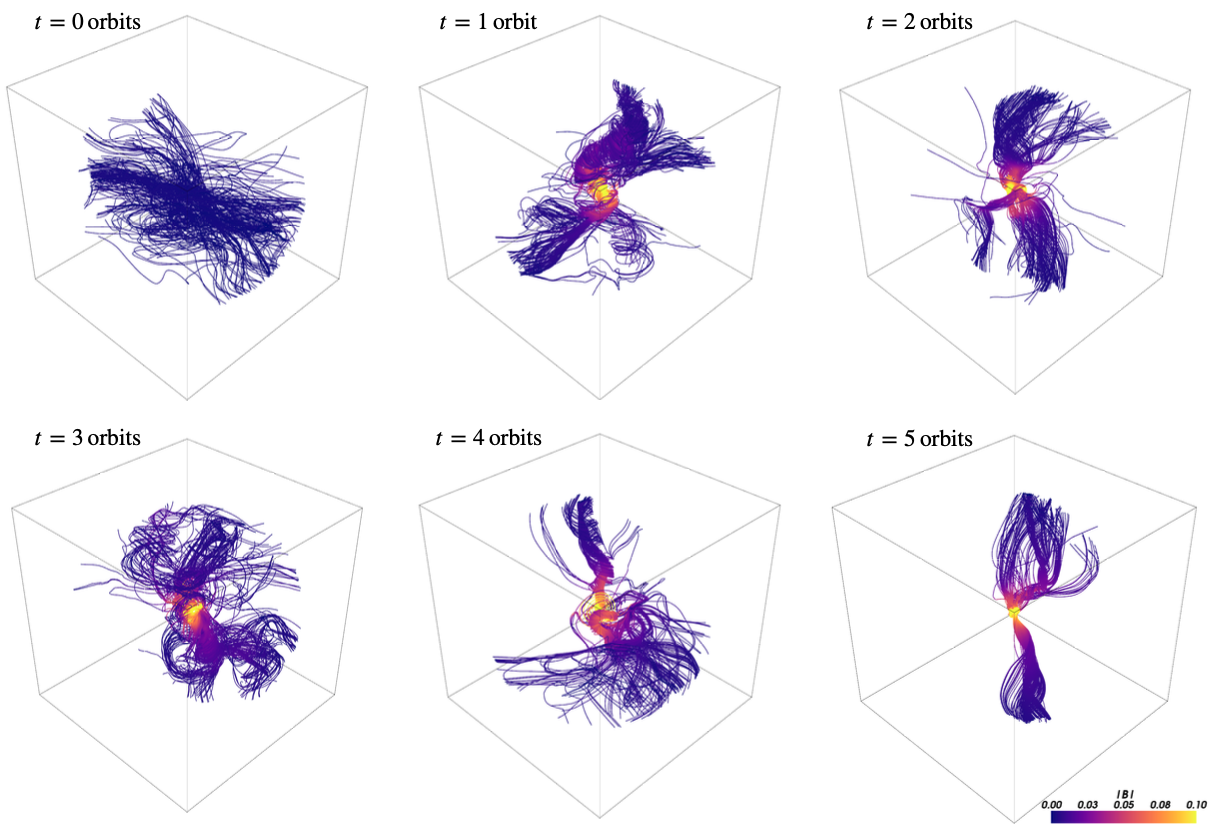}
  \caption{Temporal evolution of the torsion of the magnetic field lines in the bipolar-jet formation region for the MAG\_R1 model. At $t=0$ orbits, prior to planet insertion, the field lines show a twist perpendicular to the $z$-axis. At this time, the MRI has fully developed in the protoplanetary disk. The colorbar 
  in the lower-right corner indicates the magnetic field magnitude.}
 \label{fig:lines}
\end{figure*}

\subsection{Code and boundary conditions}

To solve the MHD Equations~(\ref{eq:gas_cont})--(\ref{eq:induc}) numerically, we use the publicly available multifluid HD/MHD code \textsc{Fargo3D} \citep[][]{BLl2019}, an extended version of the original \textsc{Fargo3D} code \citep[][]{BLlM2016} that incorporates the Rapid Advection Algorithm on Arbitrary Meshes \citep[RAM;][]{Bll2023}. However, RAM is not implemented in the MHD module. We therefore use the standard transport scheme together with the mesh-density functions in the azimuthal direction.

We implement boundary conditions for the gas density and velocity components similar to those used in \citet[][]{Chametla_Masset2021}. Specifically, at the radial boundaries, we extrapolate the density and azimuthal velocity from the active cells into the ghost cells according to the initial power-law density profile and the initial equilibrium velocity profile, respectively. For the polar velocity component, we copy the values from the active cells into the ghost cells, while for the radial velocity component we impose reflective boundary conditions. Analogous boundary conditions are applied to the magnetic-field components. Additionally, we use wave-damping zones at the radial boundaries, as in \citet{dVal2006}. The inner damping ring extends out to $r_i=0.63r_p$, whereas the outer damping ring begins at $r_o=1.59r_p$. In all models, the damping timescale at the edge of each ring is $0.3T_\mathrm{loc}$, where $T_\mathrm{loc}$ is the local orbital period. We include magnetic-resistivity buffer zones only at the radial boundaries, covering the same radial extent as the damping zones for the hydrodynamic quantities. The radial extent of the domain is chosen to minimize the influence of boundary disturbances on the bipolar jet-launching region.

At the upper and lower boundaries in colatitude, we extrapolate the density and azimuthal velocity into the ghost cells according to the vertical hydrostatic and rotational equilibrium profiles, respectively. For the radial and polar velocity components, as well as for all magnetic-field components, we copy the values from the active cells into the ghost cells. At the radial and polar boundaries, we linearly extrapolate the electromotive forces from the active cells into the ghost cells.

\subsection{Turbulent protoplanetary disk}\label{sec:turbu}

In \citet{Chametla2026}, we investigated the formation and properties of a CPD with the planetary potential included from the outset, when the disk was still laminar and threaded by an initially vertical magnetic field. In this paper, however, we allow the protoplanetary disk to evolve into a turbulent state before switching on the planetary potential. To distinguish this phase from the subsequent evolution, we use the time variable $t'$ for the disk evolution before the planet is introduced and set $t=0$ at the moment when the planet's gravitational potential is switched on. At $t'=0$, we add perturbations to the $v_r$ and $v_\theta$ components with a characteristic amplitude equal to $5\%$ of the sound speed $c_{s}$. We then evolve the disk for eight orbits, more than twice the estimated MRI growth time of approximately three orbital periods. Figure~\ref{fig:disk} shows the magnetic-field magnitude at $t'=8$ orbits, illustrating the resulting MRI-driven turbulent state.

The upper panel of Figure~\ref{fig:ic} shows the radial profile of the stress-to-pressure ratio $\alpha$ at $t'=8$ orbits, while the middle panel shows the temporal evolution of its radial average. This ratio measures the strength of the turbulent stresses relative to the gas pressure and is defined as
\begin{equation}
\alpha\left(r,t\right)=\alpha_\mathrm{Rey}\left(r,t\right)+\alpha_\mathrm{Max}\left(r,t\right)   \label{eq:alpha}
\end{equation}
with the mass-averaged Reynolds tensor contribution $\alpha_\mathrm{Rey}$ 
\begin{equation}
\alpha_\mathrm{Rey}(r,t)=\frac{1}{c_{s}^{2}(r)}\frac{\int d\Omega\,\rho \,\delta v_r\,\delta v_\phi}{\int d\Omega\, \rho },
    \label{eq:alphaR}
\end{equation}
where $d\Omega=\sin{\theta}\, d\theta\,d\phi$, $\delta v_r= v_r-\bar{v}_r$ and $\delta v_\phi= v_\phi-\bar{v}_\phi$, and
the mass-averaged Maxwell tensor contribution $\alpha_\mathrm{Max}$
\begin{equation}
\alpha_\mathrm{Max} (r,t)=-\frac{1}{\mu_0 c_{s}^{2}(r)}\frac{\int  
 d\Omega\,B_{r,c} B_{\phi,c}}{\int d\Omega\, \rho }. \label{eq:alphaM}
\end{equation}
For the model with an initial $\beta=875$, the stress parameter reaches $\alpha\simeq0.2$ at $t'=3$ orbits and remains near this level until the planet is inserted at $t'=8$ orbits. This value is comparable to those obtained in local ideal-MHD models with net vertical magnetic flux by \citet{HG1995}.

The lower panel of Figure~\ref{fig:ic} shows the product $\langle\alpha\rangle\langle\beta\rangle$, where the angle brackets denote volume averages. At late times, this product remains above the value $\approx 0.5$ commonly found in local shearing-box simulations and increases systematically with the initial magnetic-field strength. A similar relation between $\alpha$ and $\beta$ has been reported in global unstratified models by \citet{Sor2012}. However, because $\alpha$ includes both Maxwell and Reynolds contributions and $\langle\alpha\rangle\langle\beta\rangle$ is the product of two separate volume averages, this quantity cannot be directly identified with the magnetic-field correlation $-2B_{r,c}B_{\phi,c}/B^2$. The present diagnostic therefore indicates a systematic dependence of the normalized total stress on the initial magnetic-field strength, but does not by itself establish a sharp transition at $\beta\approx1000$.

\section{Results} \label{sec:results}

In this section, we focus on results from the MAG\_R1 simulation (see Table~\ref{tab:t1}), while results from the other runs are discussed in Section~\ref{sec:discussion}. In all main simulations, the gravitational potential of the planet, with $M_p=1\,M_\mathrm{Jup}$, is switched on instantaneously at $t'=8$ orbits, which defines $t=0$ (see the Appendix for a simulation in which the potential is instead switched on at $t'=10$ orbits).
We begin by comparing the morphology of the CPD in the hydrodynamic and MHD simulations.

\subsection{CPD morphology}

Figure~\ref{fig:comparison_HM} shows snapshots of the three-dimensional isodensity surfaces within one-third of the planet's Hill radius at $t=5$ orbits for our HYDRO and MAG\_R1 models. In the HYDRO model, we find the characteristic morphology of a cold isothermal CPD, consistent with previous simulations \citep[see, e.g.,][]{Ayliffe2009,Fung2019,Krapp2024,Lega2024}. The situation differs in the MHD case. Here, the CPD is somewhat smaller and, more notably, forms in a plane that is tilted relative to the midplane of the protoplanetary disk\footnote{By contrast, in an MHD simulation with identical parameters but in which the planet is introduced before turbulence develops, the CPD remains coplanar with the protoplanetary disk \citep[see][]{Chametla2026}.}.
In addition, we note helical structures emerging from its inner region (left panel of Figure~\ref{fig:comparison_HM}). As discussed in the next subsection, these structures are associated with the formation of bipolar jets in the vicinity of the planet.

Despite the short duration of our simulations, which extend to only five planetary orbits, the general morphology of the CPD becomes established within the first few orbits and evolves more slowly toward the end of our runs (Figure~\ref{fig:jets}). Although a steady state is not reached, only modest variations in CPD size and the emergence of minor substructures are observed over the simulated interval. Our strongly magnetized CPD also appears less morphologically variable than the circumplanetary flow reported in the non-ideal MHD simulations of \citet{Gressel2013}, which exhibited stochastic accretion and substantial structural variability. However, this comparison should be treated cautiously because of the different physical assumptions adopted in the simulations. In addition, the adiabatic non-magnetized simulations of circumplanetary gas by \citet{Fung2019} produced pressure-supported envelopes rather than rotationally supported disks. Within the limited duration of our MHD simulations, a coherent global CPD morphology develops in the turbulent magnetized environment, although longer simulations are required to determine its persistence.

Another notable difference in the magnetized model is the reduction of the CPD size by approximately $30\%$ relative to the HYDRO case, as shown in the top panel of Figure~\ref{fig:mt}. We define the CPD radius $R_\mathrm{CPD}$ as the location where the density decreases to one-third of its central value. In the HYDRO model, this radius coincides with the location in the CPD midplane where the specific angular momentum exhibits a sharp decline, supporting the use of this density threshold as a proxy for the CPD edge. We apply the same criterion to both models.

As shown in Figure~\ref{fig:mt}, the total CPD mass in both models reaches approximately $0.1\,M_\mathrm{Jup}$, although the CPD in the MAG\_R1 model is slightly less massive than that in the HYDRO model, consistent with its smaller physical extent. These relatively large masses motivate an assessment of the susceptibility of the CPDs to gravitational instability.
Therefore, we computed the Toomre parameter defined as 
\begin{equation}
Q_B=\frac{\kappa\sqrt{c_s^2+v_{A}^2}}{\pi G \Sigma},
    \label{eq:Toom}
\end{equation}
where $v_A$ is the Alfvén speed and $\kappa$ denotes the epicyclic frequency associated with the planet's gravitational potential
\citep[e.g.,][]{Kubli2023}. Figure~\ref{fig:Toomre} shows that $Q_B>1$ throughout both models, indicating stability against local axisymmetric gravitational instability according to this criterion. Moreover, $Q_B>5$ over most of the CPD, suggesting that gas self-gravity is unlikely to significantly affect its bulk structure and supporting its neglect in our simulations.

\subsection{Twisted bipolar-jet formation}

Another prominent feature of the magnetized turbulent flow, in addition to the large global tilt of the CPD plane described above, is the formation of bipolar jets launched approximately along the CPD rotation axis. In this section, we examine this phenomenon in detail.

In the left panels of Figure~\ref{fig:jets}, we show the temporal evolution of these twisted bipolar jets. Owing to the tilted configuration of the CPD, the jets are launched obliquely relative to the vertical axis of the protoplanetary disk and subsequently develop curved, helical structures. Their morphology varies from one orbital period to the next, likely reflecting the turbulent environment and the magnetic stresses associated with the twisted field lines. Nevertheless, their overall bipolar structure persists throughout the simulated interval. We find no evidence over this interval that the jets are quenched by gas infalling from the protoplanetary disk. As the jets curve toward directions nearly parallel to the protoplanetary disk midplane, their direct interaction with the predominantly vertical inflow may be reduced. Their persistent bipolar morphology also contrasts with the transient one-sided jet reported by \citet{Gressel2013}.

\begin{figure}
 \centering
 \includegraphics[width=1.0\linewidth]{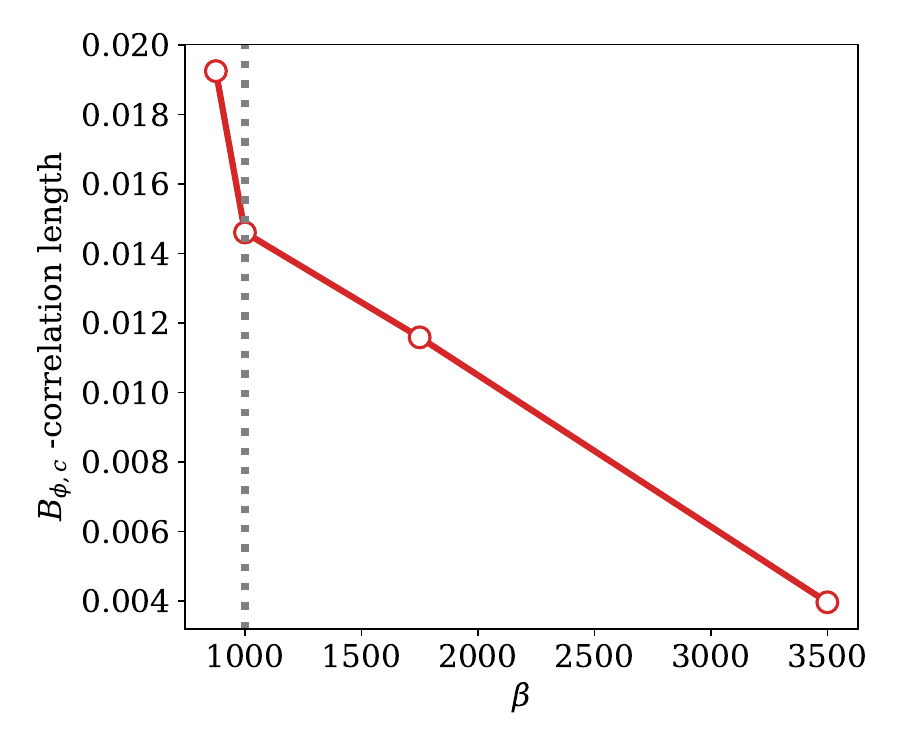}
  \caption{Spatial correlation length of the azimuthal component of the magnetic field $B_{\phi,c}$ at $t=0$
  as a function of the plasma parameter $\beta$ for the four models from MAG\_R1 to MAG\_R4.}
 \label{fig:ccross}
\end{figure}
 
\begin{figure*}
 \centering
 \includegraphics[width=1.0\linewidth]{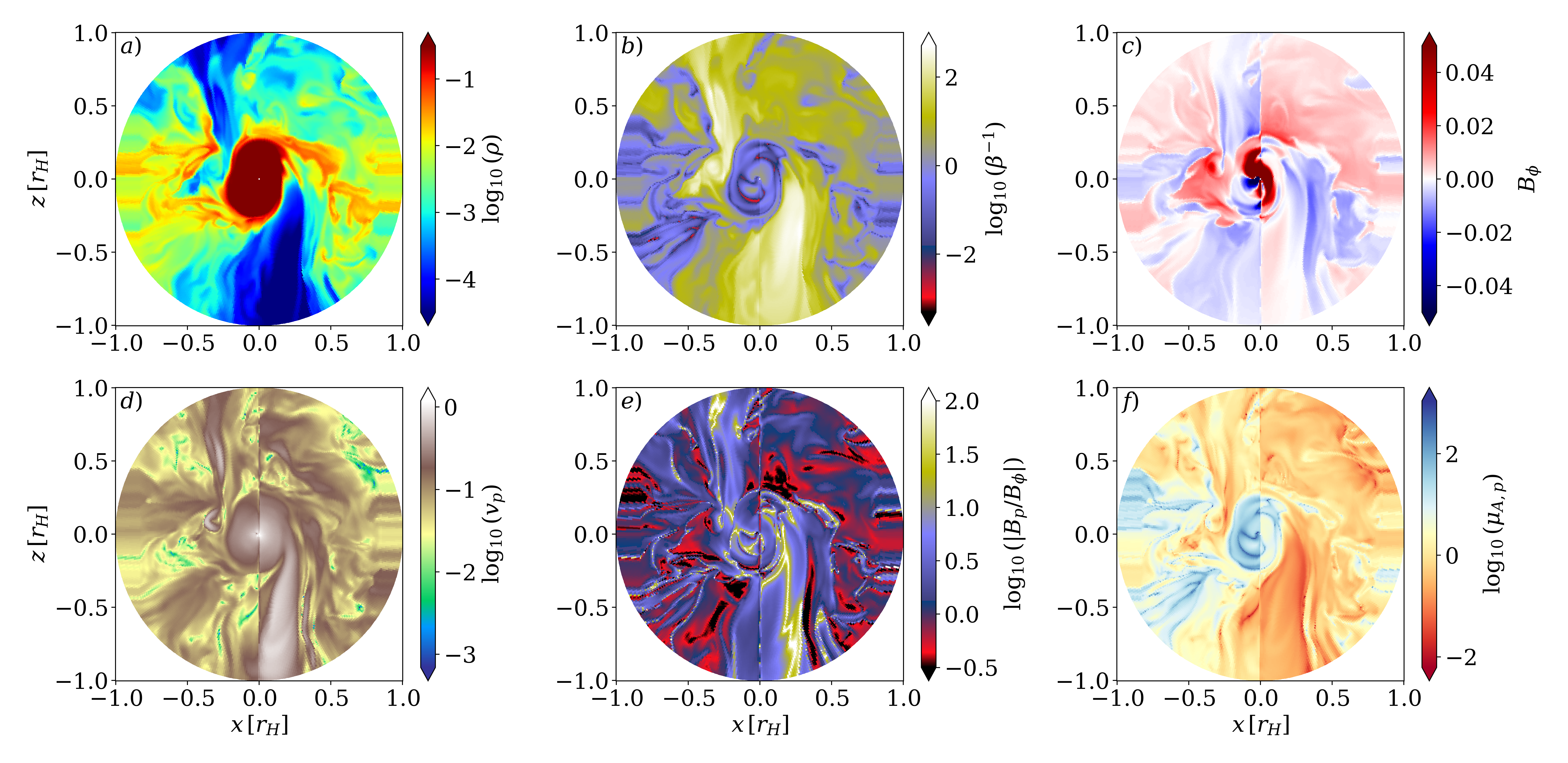}
  \caption{2D maps for the MAG\_R1 model in the $x-z$ plane, centered on the planet, showing the density $\rho$, the inverse of the beta parameter ($\beta^{-1}$), the toroidal component of the magnetic field ($B_\phi$), the poloidal component of the velocity ($v_p$), the poloidal-to-toroidal component ratio of the magnetic field and the poloidal Alfven Mach number, within the planet's Hill sphere at $t=5$ orbits. The poloidal and toroidal components are defined with respect to the planet.}
 \label{fig:fields}
\end{figure*}

Figure~\ref{fig:midplane} shows the gas-density distribution in the protoplanetary disk midplane within the planet's Hill radius for both the hydrodynamic and hydromagnetic models. In the hydrodynamic model, the familiar CPD structure occupies the region within a planetocentric distance of $0.5\,r_\mathrm{H}$, with spiral arms extending from its outer edge. By contrast, the hydromagnetic model is distinguished by the presence of (i) two high-density lobes and (ii) two smaller lobes of very low density (the dark-blue regions in the top panel of Figure~\ref{fig:midplane}), all located within the same region. Cross-correlating these features with the 3D density isosurfaces shown in Figure~\ref{fig:comparison_HM}, we find that  the high-density lobes correspond to sections of the tilted CPD viewed edge-on.

The low-density lobes are associated with the bases of the bipolar jets. This connection is evident from their locations in the $t=4\,T_0$ panel of Figure~\ref{fig:jets}. In particular, the low-density lobe around $(X,Y)=(1.005,0.005)$\footnote{Throughout the text, we use $(X,Y)$ for coordinates in the reference system anchored to the central star and $(x,y)$ for Cartesian planetocentric coordinates.} coincides with the base of the jet directed below the protoplanetary disk midplane, whereas the low-density lobe near $(X,Y)=(0.995,-0.005)$ coincides with the base of the jet directed above it. The spatial correspondence between the low-density features in Figures~\ref{fig:jets} and~\ref{fig:midplane} further supports this identification.

\section{Discussion}\label{sec:discussion}

\subsection{Tilted CPD driven by the global toroidal magnetic field development}
\label{subsec:tilted}

In the hydrodynamic model, the CPD inclination remains approximately zero, with its rotation axis aligned with the $z$-axis of the protoplanetary disk. By contrast, stochastic accretion from a quasi-turbulent circumplanetary flow has been suggested to generate CPD tilts of up to approximately $15^\circ$ \citep[see][]{Gressel2013,Mar2021}. In our MRI-driven turbulent protoplanetary disk, the CPD inclination reaches more than $80^\circ$ in the MAG\_R1 simulation, making the CPD plane nearly perpendicular to the protoplanetary disk midplane (Figure~\ref{fig:tangle}). Although the tilt angle subsequently decreases, it remains substantial over the simulated interval. Such a large misalignment could provide the initial conditions for further evolution through the tilt instability or Lidov--Kozai oscillations, provided that the CPD satisfies the conditions required for these mechanisms to operate \citep[][]{Mar2020,Mar2021,MarA2021}.

The bottom panel of Figure~\ref{fig:tangle} shows the temporal evolution of the vertical component of the specific angular momentum, defined locally in planetocentric coordinates $(r_{pc},\theta_{pc},\phi_{pc})$ as
\begin{equation}
\ell = r_{pc} v_{\phi,pc}\sin\theta_{pc}
    \label{eq:ell}
\end{equation}
and normalized by the corresponding reference value based on the Keplerian speed in the softened planetary potential,
\begin{equation}
\ell_\mathrm{Kep} = \frac{\sqrt{GM_p}\,r_{pc}^2}{(r_{pc}^2+r_\mathrm{sm}^2)^{3/4}}\sin\theta_{pc}.
    \label{eq:ell_k}
\end{equation}
The normalized vertical component decreases rapidly after the first orbital period, showing that rotation about the planetocentric $z$-axis, and hence rotational support in the $x$-$y$ plane, weakens as the CPD tilts. Because this diagnostic is a projection onto the fixed $z$-axis, however, its decrease does not necessarily imply a comparable loss of total specific angular momentum or of rotational support about the tilted CPD axis. 

We next examine whether the CPD tilt varies with the initial value of $\beta$. Figure~\ref{fig:new_models} shows that the tilt remains small in all the models explored with $\beta\gtrsim 1000$. The density distributions of these models also appear qualitatively similar upon visual inspection.

To distinguish the role of pre-existing global MRI turbulence from that of turbulence generated locally during CPD formation, we performed a simulation using the same model as that described in Section~\ref{sec:equations}, but with two modifications: the radial and polar components of the gas velocity are initially unperturbed (i.e., $\delta v_r=\delta v_\theta=0$), and the planet is inserted at $t'=0$ \citep{Chametla2026}. In this setup, the CPD begins to form before the MRI is fully developed in the protoplanetary disk, while the disk is still threaded by the initially vertical magnetic field. The simulation nevertheless produces a turbulent CPD and collimated bipolar jets, while the CPD remains aligned with the protoplanetary disk midplane. This comparison shows that locally generated turbulence and jet launching are not sufficient, in this setup, to produce a large CPD tilt. Rather, a large tilt is found only when the planet is inserted into a disk with pre-existing global MRI turbulence. These results therefore suggest that the pre-existing turbulent state, or its associated velocity and magnetic-field structure, is required under the conditions explored here.

One possible way in which this pre-existing state could tilt the CPD is through a magnetic torque. Before the planet is inserted, differential rotation and the MRI have already amplified the toroidal component of the magnetic field in the global disk. The CPD therefore forms in a region where the magnetic field is no longer purely vertical but possesses a substantial azimuthal component. The configuration shown in the first panel of Figure~\ref{fig:lines} suggests that gradients in the toroidal magnetic pressure, together with the magnetic tension associated with the curved field lines, produce an antisymmetric vertical Lorentz force: upward on the outer side of the planet's orbit ($r>r_p$) and downward on its inner side ($r<r_p$). Such a force distribution would exert a torque about an axis in the protoplanetary disk midplane and could thereby tilt the forming CPD. The resulting magnetic-field geometry is qualitatively reminiscent of the spiral field and outflow structures obtained in simulations of collapsing magnetized turbulent cloud cores \citep[see][and references therein]{Matsu2011}, although the physical setting is substantially different. Establishing whether this magnetic torque is responsible for the tilt requires a direct angular-momentum budget that also accounts for the accretion of misaligned angular momentum.
 
In addition to the volume density, Figure~\ref{fig:new_models} shows the azimuthal magnetic field $B_{\phi,c}$ (defined with
respect to the star) in the meridian $x-z$ plane for models MAG\_R1, MAG\_R2, MAG\_R3 and MAG\_R4 (see Table~\ref{tab:t1}). For these last three models, the configuration of the azimuthal magnetic field exhibits a change of sign in $X$ and $Z$, originating from the planet's position. In contrast, the MAG\_R1 model exhibits a strong winding of the azimuthal magnetic field around the planet and in the region of bipolar jet formation. Note that this azimuthal magnetic field winding has its central engine in the CPD, as can be seen in the density map. In the other models, the azimuthal magnetic field configuration shapes the jets and generates filament formation emerging from the coplanar CPDs.

The above considerations may be partially related to the fact that the spatial
structure of the azimuthal magnetic field depends on the initial value of $\beta$. To verify this, we computed the correlation length of $B_{\phi,c}$ at $t=0$ as the lag 
at which the autocorrelation function $B_{\phi,c}^{\mathrm{cross}}$ drops  
to $1/e$ of its maximum. Specifically, the autocorrelation function is given by
\begin{equation}
    B_{\phi.c}^{\mathrm{cross}}[k]=\sum_{l=0}^{N-1}B_{\phi,c}^l  B_{\phi,c}^{l-k}
    \label{eq:Bpcc}
\end{equation}
for $k=-(N-1),...,(N-1)$, with $N$ the length of $B_{\phi,c}$ and taking into account 
that $B_{\phi,c}^m=0$ if $m\notin[0,M-1]$.
As shown in Figure~\ref{fig:ccross}, the correlation length increases
markedly for $\beta<1000$.

To assess whether the CPD obliquity is sensitive to the precise state of
the MRI turbulence, we present in the Appendix the evolution of the system 
when the planet is introduced at $t'=10$ orbits. We find that the resulting tilt angle
is similar.

Hydrodynamic simulations of circumstellar disk formation in turbulent molecular cloud cores have shown that the disk angular momentum can become misaligned with the total angular momentum of the host cloud core \citep{Tsukamoto2013}. Simulations that also include magnetic fields have found misalignments among the disk angular-momentum, magnetic-field, outflow, and envelope directions \citep{Matsumoto2017}. In both studies, the turbulence is introduced through an initially imposed velocity field with a prescribed power spectrum and is not continuously driven, in contrast to the MRI-driven turbulence in our protoplanetary disk models. Despite this fundamental difference, these results demonstrate that the local angular momentum delivered to a forming disk need not be aligned with either the large-scale magnetic field or the global angular momentum. Turbulent accretion of misaligned angular momentum may therefore contribute to the CPD tilt, either independently of or in addition to a direct magnetic torque.

\begin{figure}
 \centering
 \includegraphics[width=1.0\linewidth]{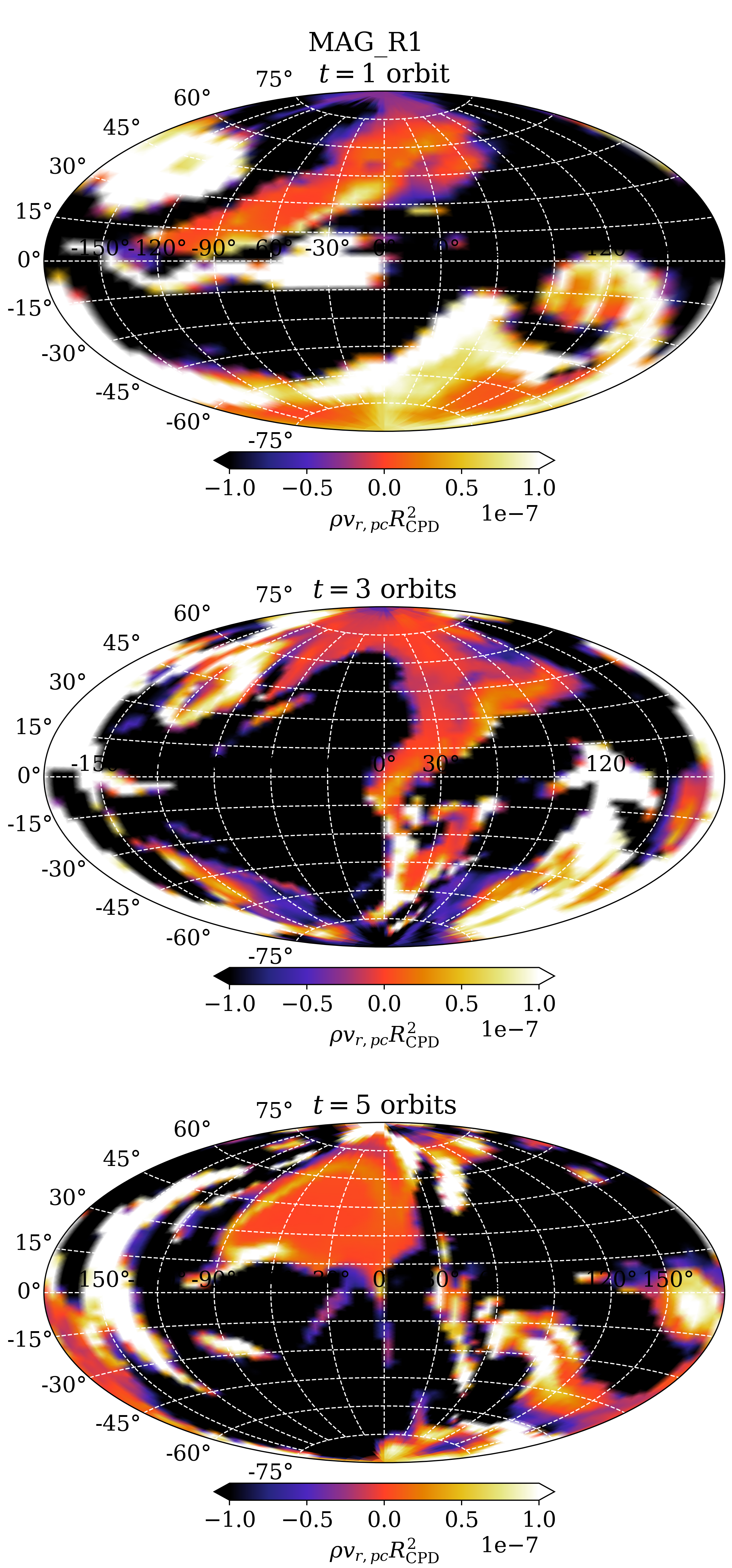}
  \caption{Mass flux through the sphere of radius $R_\mathrm{CPD}$ centered at the planet position for the MAG\_R1 model. 
  Latitude measured with respect to the protoplanetary disk midplane. Here $\rho v_{r,pc}R_\mathrm{CPD}^2>0$ indicates an inflow.}
 \label{fig:flux}
\end{figure}

\begin{figure}[t]
 \centering
 \includegraphics[width=1.0\linewidth]{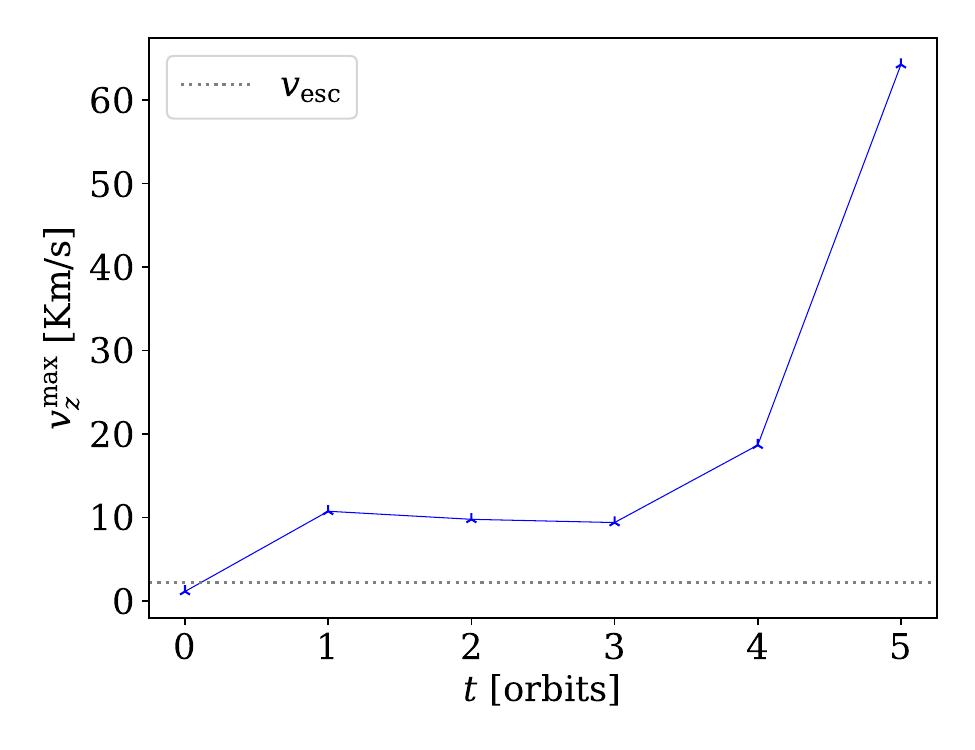}
  \caption{Temporal evolution of the maximum value of the vertical velocity $v_z^\mathrm{max}$ in the jet formation region for the hydromagnetic MAG\_R1 model. The gray dotted line represents the escape velocity of the planet.}
 \label{fig:vz}
\end{figure}

\subsection{Twisted bipolar jet ejection mechanisms}
\label{subsec:twi_jets}

Two commonly discussed magnetic mechanisms for launching and accelerating accretion-disk winds and jets are magnetocentrifugal acceleration along inclined poloidal field lines \citep[][]{BP1982,E1997} and acceleration by gradients in the toroidal magnetic pressure \citep[][]{Fukue1990,C1995}. These mechanisms are not mutually exclusive and may operate in different regions of the same outflow. Their relative importance depends on the magnetic-field geometry, disk rotation, and mass loading. The poloidal-to-toroidal field-strength ratio, $B_p/B_\phi$, defined relative to the local disk or jet axis, provides a useful diagnostic, although distinguishing the two mechanisms ultimately requires an analysis of the forces acting along the outflow.

In our MAG\_R1 simulation, the CPD is highly tilted and embedded within a turbulent protoplanetary disk, complicating the direct application of standard axisymmetric jet-launching diagnostics. As a first step toward characterizing the launching and subsequent twisting of the jets, we decompose the velocity and magnetic fields into poloidal and toroidal components in a planetocentric coordinate system aligned with the instantaneous angular-momentum vector of the CPD.

In Figure~\ref{fig:fields}, we show 2D slices in the $x$-$z$ plane of the density $\rho$, the inverse plasma-beta parameter $\beta^{-1}$, the toroidal magnetic-field component $B_\phi$, the poloidal velocity $v_p$, the poloidal-to-toroidal magnetic-field ratio $B_p/B_\phi$, and the poloidal Alfvén Mach number $\mu_{A,p}=v_p\sqrt{\mu_0\rho}/B_p$. The slices cover the planet's Hill sphere at $t=5\,T_0$. The poloidal and toroidal components are defined in a planet-centered frame whose $z$-axis is perpendicular to the protoplanetary disk midplane.
Panel~(a) shows the low-density regions associated with the twisted bipolar jets, whose bases lie close to the CPD. Panel~(b) shows that $\beta^{-1}>1$ in much of these regions, indicating that the magnetic pressure exceeds the thermal gas pressure. Panels~(c) and~(d) show a strong, spatially structured toroidal magnetic field and enhanced poloidal velocities within the CPD and the jets. The toroidal component changes sign near the narrow jet cores. Correspondingly, panel~(e) shows large values of $B_p/B_\phi$ in these regions. Because $B_\phi$ crosses zero there, however, the large ratio cannot by itself be interpreted as evidence of poloidal-field dominance or magnetocentrifugal launching. In the surrounding regions, $B_p$ and $|B_\phi|$ become comparable. Panel~(f) shows the corresponding poloidal Alfvén Mach number, which characterizes the flow speed relative to the poloidal Alfvén speed but does not independently identify the acceleration mechanism. Because this decomposition is aligned with the protoplanetary disk rather than with the highly tilted CPD, these quantities provide only a descriptive view of the magnetic and velocity structure. Distinguishing magnetocentrifugal acceleration from acceleration by toroidal magnetic-pressure gradients requires a CPD-aligned decomposition and a direct analysis of the forces along the outflow.

Figure~\ref{fig:lines} shows a sequence of magnetic-field configurations in the bipolar-jet formation region. The field lines exhibit a helical structure within the narrow jet cores, indicating substantial magnetic winding. This geometry alone, however, does not distinguish magnetocentrifugal launching from acceleration by gradients in the toroidal magnetic pressure. Broader and more weakly wound field-line configurations are also present around the jet cores. The overall morphology qualitatively resembles the spiral-flow outflows found in simulations of collapsing magnetized turbulent cloud cores \citep[see][]{Matsu2011,Matsumoto2017}. We therefore regard a spiral-flow interpretation as plausible, although identifying the dominant acceleration mechanism requires a direct analysis of the forces acting along the outflow.

The density distribution in panel~(a) of Figure~\ref{fig:fields} and the mass-flux maps in Figure~\ref{fig:flux} show that the outflow crosses the sphere of radius $R_\mathrm{CPD}$ primarily at nonpolar latitudes measured relative to the protoplanetary disk midplane. Near the poles defined by this axis, the mass flux is weak or directed inward. This pattern is consistent with the inclined and curved geometry of the jets, but it does not imply the absence of a bipolar outflow relative to the CPD axis. At none of the times shown in Figure~\ref{fig:flux} is outward mass flux present simultaneously at both protoplanetary-disk poles. The figure also shows that, from $t=3\,T_0$ onward, the inward mass flux weakens over a range of latitudes extending from the equator ($0^\circ$) toward the south pole ($-90^\circ$). This change coincides with the slowing of the CPD's radial growth shown in Figure~\ref{fig:mt}. Establishing a causal connection, however, requires evaluating the net inflow integrated over the full sphere.

\subsection{Beyond the ideal MHD approximation} 

The primary objective of this study is to analyze the formation and evolution of CPDs and its associated bipolar jets within a turbulent environment driven by the MRI. Consequently, our MHD models (MAG\_R1--MAG\_R4) omit non-ideal MHD effects, such as Ohmic resistivity and ambipolar diffusion, in the vicinity of the planet. These non-ideal processes govern the coupling between the magnetic field and the gas in high- and low-density regimes, both of which can occur during CPD evolution and within the jet launching regions, respectively. Including these effects could potentially alter the structural properties of the CPD and the orientation or collimation of the twisted bipolar jets presented here.
A comprehensive exploration of how each non-ideal MHD effect independently impacts CPD orientation and jet dynamics warrants a dedicated future study. Nevertheless, the $\alpha$-value for our MAG\_R1 model prior to planet insertion is consistent with the values found in non-ideal MHD simulations with a strong-coupling ambipolar diffusion coefficient $\mathrm{Am} \gg 1$, and an initial plasma parameter $\beta < 1000$ \citep{BS2011}. 

\section{Conclusions} \label{sec:conclusions}

We performed three-dimensional, high-resolution hydrodynamic and magnetohydrodynamic simulations of the gas flow around a Jupiter-class planet embedded in a protoplanetary disk. In the magnetohydrodynamic models, the disk is allowed to develop MRI-driven turbulence before the planet is inserted. We examined the formation, structure, and orientation of the CPD, as well as the formation of bipolar outflows. Within the parameter range explored, a large CPD tilt develops only in the most strongly magnetized model, MAG\_R1, which reaches a maximum inclination of approximately $87^\circ$, whereas the models with initial $\beta\gtrsim1000$ remain nearly coplanar. The CPD in MAG\_R1 forms already inclined rather than evolving from an initially coplanar configuration (see Figs.~\ref{fig:jets} and~\ref{fig:tangle}). Its formation is associated with the pre-existing global MRI-turbulent state and the strong azimuthal magnetic field present when the planet is inserted. The current diagnostics, however, do not distinguish whether the tilt is produced primarily by a direct magnetic torque or by the accretion of misaligned angular momentum.

We also find that twisted bipolar outflows emerge from the polar regions of the tilted CPD. The magnetic-field and flow morphology qualitatively resembles the spiral-flow outflows found in simulations of circumstellar-disk formation in magnetized turbulent cloud cores. However, the present diagnostics do not uniquely distinguish this morphology from magnetocentrifugal acceleration or acceleration by toroidal magnetic-pressure gradients. The maximum $z$-component of the velocity, measured in the protoplanetary-disk frame near the Hill radius, reaches approximately $60\,\mathrm{km\,s^{-1}}$ in the low-density jet material in some runs (see Fig.~\ref{fig:vz}). This is about three times the stellar escape velocity at the planet's orbital radius. Establishing whether this gas escapes the system nevertheless requires an analysis of its total energy in the combined star--planet potential. Such high velocities may produce observable signatures, although their detectability will also depend on the jet density, mass-loss rate, thermodynamic state, and emission properties.

However, it should be noted that a vertical velocity such as the one shown in Figure~\ref{fig:vz} is above the range of jet velocities detected in observed circumplanetary disks around massive planet candidates \citep[see][and references therein]{Dutrey2024,Zakamska2025}. Observationally, the magnetic twisting found in our MAG\_R1 model provides specific predictions for shock and kinematic tracers. The non-axisymmetric internal shocks produced by the helical instability will generate localized peaks in SO emission, mapping the path of the twisted magnetic field lines. Simultaneously, the rotation and helical motion of the collimated gas will imprint an oscillating velocity gradient on the Na D line profiles. Because the strong toroidal component ensures tight jet collimation, the shock energy and kinematic signatures remain highly concentrated rather than diluted, making these magnetic twisting features prime, distinct targets for observational campaigns searching for massive Jupiter-like planets.

Finally, a highly inclined CPD could, in principle, alter the planetary spin through gravitational or magnetic coupling and through the accretion of misaligned angular momentum. Whether this process can produce a substantial planetary obliquity depends on the relative angular momenta of the planet and CPD and on their coupling and evolutionary timescales. Our simulations do not include the planetary spin, so their applicability to the obliquities of Saturn or Uranus remains speculative. We defer a quantitative investigation of CPD--planet spin coupling to future work.

\begin{acknowledgements}
The authors thank the reviewer for their useful comments and suggestions.
R.O.C thanks Alicia Moranchel-Basurto, Gennaro D'Angelo, Yasuhiro Hasegawa and Gabriel Dominique-Marleau for useful discussions. The work of R.O.C., O.C. and D.V. was supported by the Czech Science Foundation (grant 25-16507S). 
The research leading to this work received funding from the Independent Research Fund Denmark via grant ID 10.46540/3103-00205B. The work presented here is supported by the Carlsberg Foundation, grant CF25-1297.
Computational resources were available thanks to the Ministry of Education, Youth and Sports of the Czech Republic through the e-INFRA CZ (ID:90254).
\end{acknowledgements}

\bibliographystyle{aa} 
\bibliography{example} %

@ARTICLE{Ayliffe2009,
       author = {{Ayliffe}, Ben A. and {Bate}, Matthew R.},
        title = "{Circumplanetary disc properties obtained from radiation hydrodynamical simulations of gas accretion by protoplanets}",
      journal = {\mnras},
         year = 2009,
        month = aug,
       volume = {397},
       number = {2},
        pages = {657-665},
          doi = {10.1111/j.1365-2966.2009.15002.x},
archivePrefix = {arXiv},
       eprint = {0904.4884},
 primaryClass = {astro-ph.EP},
       adsurl = {https://ui.adsabs.harvard.edu/abs/2009MNRAS.397..657A}
}

@ARTICLE{BH1991,
       author = {{Balbus}, Steven A. and {Hawley}, John F.},
        title = "{A Powerful Local Shear Instability in Weakly Magnetized Disks. I. Linear Analysis}",
      journal = {\apj},
         year = 1991,
        month = jul,
       volume = {376},
        pages = {214},
          doi = {10.1086/170270},
       adsurl = {https://ui.adsabs.harvard.edu/abs/1991ApJ...376..214B}
}

@ARTICLE{BH1998,
       author = {{Balbus}, Steven A. and {Hawley}, John F.},
        title = "{Instability, turbulence, and enhanced transport in accretion disks}",
      journal = {Reviews of Modern Physics},
         year = 1998,
        month = jan,
       volume = {70},
       number = {1},
        pages = {1-53},
          doi = {10.1103/RevModPhys.70.1},
       adsurl = {https://ui.adsabs.harvard.edu/abs/1998RvMP...70....1B}
}

@ARTICLE{BatM2020ApJ,
       author = {{Batygin}, Konstantin and {Morbidelli}, Alessandro},
        title = "{Formation of Giant Planet Satellites}",
      journal = {\apj},
         year = 2020,
        month = may,
       volume = {894},
       number = {2},
          eid = {143},
        pages = {143},
          doi = {10.3847/1538-4357/ab8937},
archivePrefix = {arXiv},
       eprint = {2005.08330},
 primaryClass = {astro-ph.EP},
       adsurl = {https://ui.adsabs.harvard.edu/abs/2020ApJ...894..143B}
}

@ARTICLE{BllM2016,
       author = {{Ben{\'\i}tez-Llambay}, Pablo and {Masset}, Fr{\'e}d{\'e}ric S.},
        title = "{FARGO3D: A New GPU-oriented MHD Code}",
      journal = {\apjs},
         year = 2016,
        month = mar,
       volume = {223},
       number = {1},
          eid = {11},
        pages = {11},
          doi = {10.3847/0067-0049/223/1/11},
archivePrefix = {arXiv},
       eprint = {1602.02359},
 primaryClass = {astro-ph.IM},
       adsurl = {https://ui.adsabs.harvard.edu/abs/2016ApJS..223...11B}
}

@ARTICLE{Bll2019,
       author = {{Ben{\'\i}tez-Llambay}, Pablo and {Krapp}, Leonardo and {Pessah}, Martin E.},
        title = "{Asymptotically Stable Numerical Method for Multispecies Momentum Transfer: Gas and Multifluid Dust Test Suite and Implementation in FARGO3D}",
      journal = {\apjs},
         year = 2019,
        month = apr,
       volume = {241},
       number = {2},
          eid = {25},
        pages = {25},
          doi = {10.3847/1538-4365/ab0a0e},
archivePrefix = {arXiv},
       eprint = {1811.07925},
 primaryClass = {astro-ph.EP},
       adsurl = {https://ui.adsabs.harvard.edu/abs/2019ApJS..241...25B}
}

@ARTICLE{Bll2023,
       author = {{Ben{\'\i}tez-Llambay}, Pablo and {Krapp}, Leonardo and {Ramos}, Ximena S. and {Kratter}, Kaitlin M.},
        title = "{RAM: Rapid Advection Algorithm on Arbitrary Meshes}",
      journal = {\apj},
         year = 2023,
        month = aug,
       volume = {952},
       number = {2},
          eid = {106},
        pages = {106},
          doi = {10.3847/1538-4357/acd698},
archivePrefix = {arXiv},
       eprint = {2305.05362},
 primaryClass = {astro-ph.IM},
       adsurl = {https://ui.adsabs.harvard.edu/abs/2023ApJ...952..106B}
}

@ARTICLE{Fujii2017,
       author = {{Fujii}, Yuri I. and {Kobayashi}, Hiroshi and {Takahashi}, Sanemichi Z. and {Gressel}, Oliver},
        title = "{Orbital Evolution of Moons in Weakly Accreting Circumplanetary Disks}",
      journal = {\aj},
         year = 2017,
        month = apr,
       volume = {153},
       number = {4},
          eid = {194},
        pages = {194},
          doi = {10.3847/1538-3881/aa647d},
archivePrefix = {arXiv},
       eprint = {1703.03759},
 primaryClass = {astro-ph.EP},
       adsurl = {https://ui.adsabs.harvard.edu/abs/2017AJ....153..194F}
}

@ARTICLE{Chametla2025,
       author = {{Chametla}, R.~O. and {Chrenko}, O. and {Masset}, F.~S. and {D'Angelo}, G. and {Nesvorn{\'y}}, D.},
        title = "{Dust-void evolution driven by turbulent dust flux can induce runaway migration of Earth-mass planets}",
      journal = {\aap},
         year = 2025,
        month = jun,
       volume = {698},
          eid = {A21},
        pages = {A21},
          doi = {10.1051/0004-6361/202451869},
archivePrefix = {arXiv},
       eprint = {2503.21922},
 primaryClass = {astro-ph.EP},
       adsurl = {https://ui.adsabs.harvard.edu/abs/2025A&A...698A..21C}
}

@ARTICLE{Fung2019,
       author = {{Fung}, Jeffrey and {Zhu}, Zhaohuan and {Chiang}, Eugene},
        title = "{Circumplanetary Disk Dynamics in the Isothermal and Adiabatic Limits}",
      journal = {\apj},
         year = 2019,
        month = dec,
       volume = {887},
       number = {2},
          eid = {152},
        pages = {152},
          doi = {10.3847/1538-4357/ab53da},
archivePrefix = {arXiv},
       eprint = {1909.09655},
 primaryClass = {astro-ph.EP},
       adsurl = {https://ui.adsabs.harvard.edu/abs/2019ApJ...887..152F}
}

@ARTICLE{G2002,
       author = {{D'Angelo}, G. and {Henning}, T. and {Kley}, W.},
        title = "{Nested-grid calculations of disk-planet interaction}",
      journal = {\aap},
         year = 2002,
        month = apr,
       volume = {385},
        pages = {647-670},
          doi = {10.1051/0004-6361:20020173},
archivePrefix = {arXiv},
       eprint = {astro-ph/0112429},
 primaryClass = {astro-ph},
       adsurl = {https://ui.adsabs.harvard.edu/abs/2002A&A...385..647D}
}

@ARTICLE{G2003,
       author = {{D'Angelo}, Gennaro and {Kley}, Willy and {Henning}, Thomas},
        title = "{Orbital Migration and Mass Accretion of Protoplanets in Three-dimensional Global Computations with Nested Grids}",
      journal = {\apj},
         year = 2003,
        month = mar,
       volume = {586},
       number = {1},
        pages = {540-561},
          doi = {10.1086/367555},
archivePrefix = {arXiv},
       eprint = {astro-ph/0308055},
 primaryClass = {astro-ph},
       adsurl = {https://ui.adsabs.harvard.edu/abs/2003ApJ...586..540D}
}

@ARTICLE{GW2023,
       author = {{Wafflard-Fernandez}, Gaylor and {Lesur}, Geoffroy},
        title = "{Planet-disk-wind interaction: The magnetized fate of protoplanets}",
      journal = {\aap},
         year = 2023,
        month = sep,
       volume = {677},
          eid = {A70},
        pages = {A70},
          doi = {10.1051/0004-6361/202245305},
archivePrefix = {arXiv},
       eprint = {2305.11784},
 primaryClass = {astro-ph.SR},
       adsurl = {https://ui.adsabs.harvard.edu/abs/2023A&A...677A..70W}
}

@ARTICLE{Gressel2013,
       author = {{Gressel}, O. and {Nelson}, R.~P. and {Turner}, N.~J. and {Ziegler}, U.},
        title = "{Global Hydromagnetic Simulations of a Planet Embedded in a Dead Zone: Gap Opening, Gas Accretion, and Formation of a Protoplanetary Jet}",
      journal = {\apj},
         year = 2013,
        month = dec,
       volume = {779},
       number = {1},
          eid = {59},
        pages = {59},
          doi = {10.1088/0004-637X/779/1/59},
archivePrefix = {arXiv},
       eprint = {1309.2871},
 primaryClass = {astro-ph.EP},
       adsurl = {https://ui.adsabs.harvard.edu/abs/2013ApJ...779...59G}
}

@ARTICLE{Kley1999,
       author = {{Kley}, Willy},
        title = "{The tidal interaction between planets and the protoplanetary disk}",
      journal = {arXiv e-prints},
         year = 1999,
        month = sep,
          eid = {astro-ph/9909394},
        pages = {astro-ph/9909394},
          doi = {10.48550/arXiv.astro-ph/9909394},
archivePrefix = {arXiv},
       eprint = {astro-ph/9909394},
 primaryClass = {astro-ph},
       adsurl = {https://ui.adsabs.harvard.edu/abs/1999astro.ph..9394K}
}

@ARTICLE{Krapp2024,
       author = {{Krapp}, Leonardo and {Kratter}, Kaitlin M. and {Youdin}, Andrew N. and {Ben{\'\i}tez-Llambay}, Pablo and {Masset}, Fr{\'e}d{\'e}ric and {Armitage}, Philip J.},
        title = "{A Thermodynamic Criterion for the Formation of Circumplanetary Disks}",
      journal = {\apj},
         year = 2024,
        month = oct,
       volume = {973},
       number = {2},
          eid = {153},
        pages = {153},
          doi = {10.3847/1538-4357/ad644a},
archivePrefix = {arXiv},
       eprint = {2402.14638},
 primaryClass = {astro-ph.EP},
       adsurl = {https://ui.adsabs.harvard.edu/abs/2024ApJ...973..153K}
}

@ARTICLE{Lubow1999,
       author = {{Lubow}, S.~H. and {Seibert}, M. and {Artymowicz}, P.},
        title = "{Disk Accretion onto High-Mass Planets}",
      journal = {\apj},
         year = 1999,
        month = dec,
       volume = {526},
       number = {2},
        pages = {1001-1012},
          doi = {10.1086/308045},
archivePrefix = {arXiv},
       eprint = {astro-ph/9910404},
 primaryClass = {astro-ph},
       adsurl = {https://ui.adsabs.harvard.edu/abs/1999ApJ...526.1001L}
}

@ARTICLE{LM2013,
       author = {{Lubow}, Stephen H. and {Martin}, Rebecca G.},
        title = "{Dead zones in circumplanetary discs as formation sites for regular satellites}",
      journal = {\mnras},
         year = 2013,
        month = jan,
       volume = {428},
       number = {3},
        pages = {2668-2673},
          doi = {10.1093/mnras/sts229},
archivePrefix = {arXiv},
       eprint = {1210.4579},
 primaryClass = {astro-ph.EP},
       adsurl = {https://ui.adsabs.harvard.edu/abs/2013MNRAS.428.2668L}
}

@ARTICLE{Lega2024,
       author = {{Lega}, E. and {Benisty}, M. and {Cridland}, A. and {Morbidelli}, A. and {Schulik}, M. and {Lambrechts}, M.},
        title = "{Gas dynamics around a Jupiter-mass planet: I. Influence of protoplanetary disk properties}",
      journal = {\aap},
         year = 2024,
        month = oct,
       volume = {690},
          eid = {A183},
        pages = {A183},
          doi = {10.1051/0004-6361/202450899},
archivePrefix = {arXiv},
       eprint = {2408.12233},
 primaryClass = {astro-ph.EP},
       adsurl = {https://ui.adsabs.harvard.edu/abs/2024A&A...690A.183L}
}

@ARTICLE{MB2016,
       author = {{Masset}, F.~S. and {Ben{\'\i}tez-Llambay}, P.},
        title = "{Horseshoe Drag in Three-dimensional Globally Isothermal Disks}",
      journal = {\apj},
         year = 2016,
        month = jan,
       volume = {817},
       number = {1},
          eid = {19},
        pages = {19},
          doi = {10.3847/0004-637X/817/1/19},
archivePrefix = {arXiv},
       eprint = {1511.07946},
 primaryClass = {astro-ph.EP},
       adsurl = {https://ui.adsabs.harvard.edu/abs/2016ApJ...817...19M}
}

@ARTICLE{Marleau2023,
       author = {{Marleau}, Gabriel-Dominique and {Kuiper}, Rolf and {B{\'e}thune}, William and {Mordasini}, Christoph},
        title = "{The Planetary Accretion Shock. III. Smoothing-free 2.5D Simulations and Calculation of H{\ensuremath{\alpha}} Emission}",
      journal = {\apj},
         year = 2023,
        month = jul,
       volume = {952},
       number = {1},
          eid = {89},
        pages = {89},
          doi = {10.3847/1538-4357/accf12},
archivePrefix = {arXiv},
       eprint = {2305.01679},
 primaryClass = {astro-ph.EP},
       adsurl = {https://ui.adsabs.harvard.edu/abs/2023ApJ...952...89M}
}

@ARTICLE{Mizuno1980,
       author = {{Mizuno}, H.},
        title = "{Formation of the Giant Planets}",
      journal = {Progress of Theoretical Physics},
         year = 1980,
        month = aug,
       volume = {64},
       number = {2},
        pages = {544-557},
          doi = {10.1143/PTP.64.544},
       adsurl = {https://ui.adsabs.harvard.edu/abs/1980PThPh..64..544M}
}

@ARTICLE{Pollack1996,
       author = {{Pollack}, James B. and {Hubickyj}, Olenka and {Bodenheimer}, Peter and {Lissauer}, Jack J. and {Podolak}, Morris and {Greenzweig}, Yuval},
        title = "{Formation of the Giant Planets by Concurrent Accretion of Solids and Gas}",
      journal = {\icarus},
         year = 1996,
        month = nov,
       volume = {124},
       number = {1},
        pages = {62-85},
          doi = {10.1006/icar.1996.0190},
       adsurl = {https://ui.adsabs.harvard.edu/abs/1996Icar..124...62P}
}

@ARTICLE{SS1973,
       author = {{Shakura}, N.~I. and {Sunyaev}, R.~A.},
        title = "{Black holes in binary systems. Observational appearance.}",
      journal = {\aap},
         year = 1973,
        month = jan,
       volume = {24},
        pages = {337-355},
       adsurl = {https://ui.adsabs.harvard.edu/abs/1973A&A....24..337S}
}

@ARTICLE{Szulagyi2016,
       author = {{Szul{\'a}gyi}, J. and {Masset}, F. and {Lega}, E. and {Crida}, A. and {Morbidelli}, A. and {Guillot}, T.},
        title = "{Circumplanetary disc or circumplanetary envelope?}",
      journal = {\mnras},
         year = 2016,
        month = aug,
       volume = {460},
       number = {3},
        pages = {2853-2861},
          doi = {10.1093/mnras/stw1160},
archivePrefix = {arXiv},
       eprint = {1605.04586},
 primaryClass = {astro-ph.EP},
       adsurl = {https://ui.adsabs.harvard.edu/abs/2016MNRAS.460.2853S}
}

@ARTICLE{Tsukamoto2013,
       author = {{Tsukamoto}, Yusuke and {Machida}, Masahiro N.},
        title = "{Formation and early evolution of circumstellar discs in turbulent molecular cloud cores}",
      journal = {\mnras},
         year = 2013,
        month = jan,
       volume = {428},
       number = {2},
        pages = {1321-1334},
          doi = {10.1093/mnras/sts111},
archivePrefix = {arXiv},
       eprint = {1210.0526},
 primaryClass = {astro-ph.SR},
       adsurl = {https://ui.adsabs.harvard.edu/abs/2013MNRAS.428.1321T}
}

@ARTICLE{Matsumoto2017,
       author = {{Matsumoto}, Tomoaki and {Machida}, Masahiro N. and {Inutsuka}, Shu-ichiro},
        title = "{Circumstellar Disks and Outflows in Turbulent Molecular Cloud Cores: Possible Formation Mechanism for Misaligned Systems}",
      journal = {\apj},
         year = 2017,
        month = apr,
       volume = {839},
       number = {1},
          eid = {69},
        pages = {69},
          doi = {10.3847/1538-4357/aa6a1c},
archivePrefix = {arXiv},
       eprint = {1703.09139},
 primaryClass = {astro-ph.SR},
       adsurl = {https://ui.adsabs.harvard.edu/abs/2017ApJ...839...69M}
}

@ARTICLE{Tanigawa2012,
       author = {{Tanigawa}, Takayuki and {Ohtsuki}, Keiji and {Machida}, Masahiro N.},
        title = "{Distribution of Accreting Gas and Angular Momentum onto Circumplanetary Disks}",
      journal = {\apj},
         year = 2012,
        month = mar,
       volume = {747},
       number = {1},
          eid = {47},
        pages = {47},
          doi = {10.1088/0004-637X/747/1/47},
archivePrefix = {arXiv},
       eprint = {1112.3706},
 primaryClass = {astro-ph.EP},
       adsurl = {https://ui.adsabs.harvard.edu/abs/2012ApJ...747...47T}
}

@ARTICLE{Ohashi2025,
       author = {{Ohashi}, Satoshi and {Muto}, Takayuki and {Tsukamoto}, Yusuke and {Kataoka}, Akimasa and {Tsukagoshi}, Takashi and {Momose}, Munetake and {Fukagawa}, Misato and {Sakai}, Nami},
        title = "{Observationally derived magnetic field strength and 3D components in the HD 142527 disk}",
      journal = {Nature Astronomy},
         year = 2025,
        month = apr,
       volume = {9},
        pages = {526-534},
          doi = {10.1038/s41550-024-02454-x},
archivePrefix = {arXiv},
       eprint = {2502.06030},
 primaryClass = {astro-ph.EP},
       adsurl = {https://ui.adsabs.harvard.edu/abs/2025NatAs...9..526O}
}

@ARTICLE{Fukue1990,
       author = {{Fukue}, Jun},
        title = "{Magnetohydrodynamical Winds from an Accretion Disk: Driving by the Magnetic Pressure and Tension of Toroidal Magnetic Fields}",
      journal = {\pasj},
         year = 1990,
        month = dec,
       volume = {42},
       number = {6},
        pages = {793-817},
          doi = {10.1093/pasj/42.6.793},
       adsurl = {https://ui.adsabs.harvard.edu/abs/1990PASJ...42..793F}
}

@ARTICLE{BP1982,
       author = {{Blandford}, R.~D. and {Payne}, D.~G.},
        title = "{Hydromagnetic flows from accretion disks and the production of radio jets.}",
      journal = {\mnras},
         year = 1982,
        month = jun,
       volume = {199},
        pages = {883-903},
          doi = {10.1093/mnras/199.4.883},
       adsurl = {https://ui.adsabs.harvard.edu/abs/1982MNRAS.199..883B}
}

@ARTICLE{C1995,
       author = {{Contopoulos}, J.},
        title = "{A Simple Type of Magnetically Driven Jets: an Astrophysical Plasma Gun}",
      journal = {\apj},
         year = 1995,
        month = sep,
       volume = {450},
        pages = {616},
          doi = {10.1086/176170},
       adsurl = {https://ui.adsabs.harvard.edu/abs/1995ApJ...450..616C}
}

@ARTICLE{E1997,
       author = {{Ostriker}, Eve C.},
        title = "{Self-similar Magnetocentrifugal Disk Winds with Cylindrical Asymptotics}",
      journal = {\apj},
         year = 1997,
        month = sep,
       volume = {486},
       number = {1},
        pages = {291-306},
          doi = {10.1086/304513},
archivePrefix = {arXiv},
       eprint = {astro-ph/9705226},
 primaryClass = {astro-ph},
       adsurl = {https://ui.adsabs.harvard.edu/abs/1997ApJ...486..291O}
}

@ARTICLE{Mar2020,
       author = {{Martin}, Rebecca G. and {Zhu}, Zhaohuan and {Armitage}, Philip J.},
        title = "{A Fast-growing Tilt Instability of Detached Circumplanetary Disks}",
      journal = {\apjl},
         year = 2020,
        month = jul,
       volume = {898},
       number = {1},
          eid = {L26},
        pages = {L26},
          doi = {10.3847/2041-8213/aba3c1},
archivePrefix = {arXiv},
       eprint = {2007.05022},
 primaryClass = {astro-ph.EP},
       adsurl = {https://ui.adsabs.harvard.edu/abs/2020ApJ...898L..26M}
}

@ARTICLE{Mar2021,
       author = {{Martin}, Rebecca G. and {Zhu}, Zhaohuan and {Armitage}, Philip J. and {Yang}, Chao-Chin and {Baehr}, Hans},
        title = "{Kozai-Lidov oscillations triggered by a tilt instability of detached circumplanetary discs}",
      journal = {\mnras},
         year = 2021,
        month = apr,
       volume = {502},
       number = {3},
        pages = {4426-4434},
          doi = {10.1093/mnras/stab232},
archivePrefix = {arXiv},
       eprint = {2101.09388},
 primaryClass = {astro-ph.EP},
       adsurl = {https://ui.adsabs.harvard.edu/abs/2021MNRAS.502.4426M}
}

@ARTICLE{MarA2021,
       author = {{Martin}, Rebecca G. and {Armitage}, Philip J.},
        title = "{Primordial Giant Planet Obliquity Driven by a Circumplanetary Disk}",
      journal = {\apjl},
         year = 2021,
        month = may,
       volume = {912},
       number = {1},
          eid = {L16},
        pages = {L16},
          doi = {10.3847/2041-8213/abf736},
archivePrefix = {arXiv},
       eprint = {2104.06479},
 primaryClass = {astro-ph.EP},
       adsurl = {https://ui.adsabs.harvard.edu/abs/2021ApJ...912L..16M}
}

@ARTICLE{Kubli2023,
       author = {{Kubli}, Noah and {Mayer}, Lucio and {Deng}, Hongping},
        title = "{Characterizing fragmentation and sub-Jovian clump properties in magnetized young protoplanetary discs}",
      journal = {\mnras},
         year = 2023,
        month = oct,
       volume = {525},
       number = {2},
        pages = {2731-2749},
          doi = {10.1093/mnras/stad2478},
archivePrefix = {arXiv},
       eprint = {2303.04163},
 primaryClass = {astro-ph.EP},
       adsurl = {https://ui.adsabs.harvard.edu/abs/2023MNRAS.525.2731K}
}

@ARTICLE{Matsu2011,
       author = {{Matsumoto}, Tomoyuki and {Hanawa}, Tomoaki},
        title = "{Protostellar Collapse of Magneto-turbulent Cloud Cores: Shape During Collapse and Outflow Formation}",
      journal = {\apj},
         year = 2011,
        month = feb,
       volume = {728},
       number = {1},
          eid = {47},
        pages = {47},
          doi = {10.1088/0004-637X/728/1/47},
archivePrefix = {arXiv},
       eprint = {1008.3984},
 primaryClass = {astro-ph.SR},
       adsurl = {https://ui.adsabs.harvard.edu/abs/2011ApJ...728...47M}
}

@ARTICLE{Sor2012,
       author = {{Sorathia}, Kareem A. and {Reynolds}, Christopher S. and {Stone}, James M. and {Beckwith}, Kris},
        title = "{Global Simulations of Accretion Disks. I. Convergence and Comparisons with Local Models}",
      journal = {\apj},
         year = 2012,
        month = apr,
       volume = {749},
       number = {2},
          eid = {189},
        pages = {189},
          doi = {10.1088/0004-637X/749/2/189},
archivePrefix = {arXiv},
       eprint = {1106.4019},
 primaryClass = {astro-ph.HE},
       adsurl = {https://ui.adsabs.harvard.edu/abs/2012ApJ...749..189S}
}

@ARTICLE{BS2011,
       author = {{Bai}, Xue-Ning and {Stone}, James M.},
        title = "{Effect of Ambipolar Diffusion on the Nonlinear Evolution of Magnetorotational Instability in Weakly Ionized Disks}",
      journal = {\apj},
         year = 2011,
        month = aug,
       volume = {736},
       number = {2},
          eid = {144},
        pages = {144},
          doi = {10.1088/0004-637X/736/2/144},
archivePrefix = {arXiv},
       eprint = {1103.1380},
 primaryClass = {astro-ph.EP},
       adsurl = {https://ui.adsabs.harvard.edu/abs/2011ApJ...736..144B}
}

@ARTICLE{Rav2017,
       author = {{Rafikov}, Roman R.},
        title = "{Protoplanetary Disks as (Possibly) Viscous Disks}",
      journal = {\apj},
         year = 2017,
        month = mar,
       volume = {837},
       number = {2},
          eid = {163},
        pages = {163},
          doi = {10.3847/1538-4357/aa6249},
archivePrefix = {arXiv},
       eprint = {1701.02352},
 primaryClass = {astro-ph.EP},
       adsurl = {https://ui.adsabs.harvard.edu/abs/2017ApJ...837..163R}
}

@ARTICLE{HG1995,
       author = {{Hawley}, John F. and {Gammie}, Charles F. and {Balbus}, Steven A.},
        title = "{Local Three-dimensional Magnetohydrodynamic Simulations of Accretion Disks}",
      journal = {\apj},
         year = 1995,
        month = feb,
       volume = {440},
        pages = {742},
          doi = {10.1086/175311},
       adsurl = {https://ui.adsabs.harvard.edu/abs/1995ApJ...440..742H}
}

@ARTICLE{Chametla2026,
       author = {{Chametla}, Ra{\'u}l O. and {S{\'a}nchez-Salcedo}, F.~J. and {Pessah}, Martin E. and {Reyes-Ruiz}, Mauricio},
        title = "{Global simulations of accretion flows onto perturbers embedded in magnetized disks -I. MRI and jet formation in ideal MHD}",
      journal = {arXiv e-prints},
         year = 2026,
        month = aug,
          eid = {arXiv:2608.09471},
        pages = {arXiv:2608.09471},
          doi = {10.48550/arXiv.2608.09471},
archivePrefix = {arXiv},
       eprint = {2608.09471},
 primaryClass = {astro-ph.HE},
       adsurl = {https://ui.adsabs.harvard.edu/abs/2026arXiv260809471C}
}

@ARTICLE{Chametla_Masset2021,
       author = {{Chametla}, Ra{\'u}l O. and {Masset}, Fr{\'e}d{\'e}ric S.},
        title = "{Numerical study of coorbital thermal torques on cold or hot satellites}",
      journal = {\mnras},
         year = 2021,
        month = jan,
       volume = {501},
       number = {1},
        pages = {24-35},
          doi = {10.1093/mnras/staa3681},
archivePrefix = {arXiv},
       eprint = {2011.12484},
 primaryClass = {astro-ph.EP},
       adsurl = {https://ui.adsabs.harvard.edu/abs/2021MNRAS.501...24C}
}

@ARTICLE{dVal2006,
       author = {{de Val-Borro}, M. and {Edgar}, R.~G. and {Artymowicz}, P. and {Ciecielag}, P. and {Cresswell}, P. and {D'Angelo}, G. and {Delgado-Donate}, E.~J. and {Dirksen}, G. and {Fromang}, S. and {Gawryszczak}, A. and {Klahr}, H. and {Kley}, W. and {Lyra}, W. and {Masset}, F. and {Mellema}, G. and {Nelson}, R.~P. and {Paardekooper}, S. -J. and {Peplinski}, A. and {Pierens}, A. and {Plewa}, T. and {Rice}, K. and {Sch{\"a}fer}, C. and {Speith}, R.},
        title = "{A comparative study of disc-planet interaction}",
      journal = {\mnras},
         year = 2006,
        month = aug,
       volume = {370},
       number = {2},
        pages = {529-558},
          doi = {10.1111/j.1365-2966.2006.10488.x},
archivePrefix = {arXiv},
       eprint = {astro-ph/0605237},
 primaryClass = {astro-ph},
       adsurl = {https://ui.adsabs.harvard.edu/abs/2006MNRAS.370..529D}
}

@ARTICLE{Dutrey2024,
       author = {{Dutrey}, A. and {Chapillon}, E. and {Guilloteau}, S. and {Tang}, Y.~W. and {Boccaletti}, A. and {Bouscasse}, L. and {Collin-Dufresne}, T. and {Di Folco}, E. and {Fuente}, A. and {Pi{\'e}tu}, V. and {Rivi{\`e}re-Marichalar}, P. and {Semenov}, D.},
        title = "{Sulfur monoxide (SO) as a shock tracer in protoplanetary disks: Case of AB Aurigae}",
      journal = {\aap},
         year = 2024,
        month = sep,
       volume = {689},
          eid = {L7},
        pages = {L7},
          doi = {10.1051/0004-6361/202451299},
archivePrefix = {arXiv},
       eprint = {2408.14276},
 primaryClass = {astro-ph.EP},
       adsurl = {https://ui.adsabs.harvard.edu/abs/2024A&A...689L...7D}
}

@ARTICLE{Zakamska2025,
       author = {{Zakamska}, Nadia L. and {Pallathadka}, Gautham Adamane and {Bizyaev}, Dmitry and {Merc}, Jaroslav and {Owen}, James E. and {Reggiani}, Henrique and {Schlaufman}, Kevin C. and {B{\k{a}}kowska}, Karolina and {Bednarz}, S{\l}awomir and {Bernacki}, Krzysztof and {Gurgul}, Agnieszka and {Hall}, Kirsten R. and {Hambsch}, Franz-Josef and {Joachimczyk}, Barbara and {Kotysz}, Krzysztof and {Kurowski}, Sebastian and {Liakos}, Alexios and {Miko{\l}ajczyk}, Przemys{\l}aw J. and {Pak{\v{s}}tien{\.{e}}}, Erika and {Pojma{\'n}ski}, Grzegorz and {Popowicz}, Adam and {Reichart}, Daniel E. and {Wyrzykowski}, {\L}ukasz and {Zdanavi{\v{c}}ius}, Justas and {{\.Z}ejmo}, Micha{\l} and {Zieli{\'n}ski}, Pawe{\l} and {Zola}, Staszek},
        title = "{ASASSN-24fw: Candidate Gas-rich Circumsecondary Disk Occultation of a Main-sequence Star}",
      journal = {\aj},
         year = 2026,
        month = feb,
       volume = {171},
       number = {2},
          eid = {95},
        pages = {95},
          doi = {10.3847/1538-3881/ae1fd9},
archivePrefix = {arXiv},
       eprint = {2507.05367},
 primaryClass = {astro-ph.EP},
       adsurl = {https://ui.adsabs.harvard.edu/abs/2026AJ....171...95Z}
}

\appendix
\section{Influence of MRI Saturation and Planet Insertion Timing on Jet Twisting}
\label{app:A1}

\begin{figure*}
    \centering
    \includegraphics[width=1.0\linewidth]{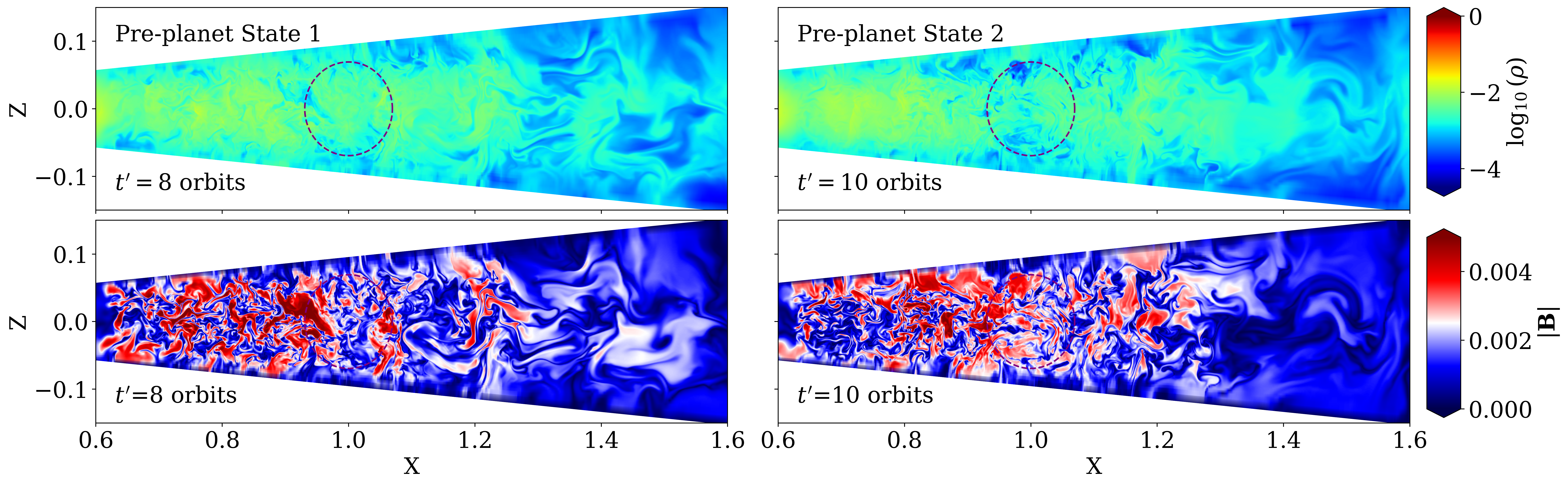}
    \caption{Vertical slices of the gas density (upper panels) and magnetic field 
    strength (lower panels) for the MAG\_R1 model at two different states $t'=8$ (left column) and $t'=10$ orbits (right column).
}
    \label{fig:IC2}
\end{figure*}

In all the experiments described in the main text, the planet was inserted at $t'=8$ orbits, after MRI-driven turbulence had fully developed throughout the global disk. To assess the robustness of our results to the precise state of the turbulent disk, we have extended the MAG\_R1 simulation to $t'=10$ orbits.
Fig. \ref{fig:IC2} shows vertical slices of density and magnetic field at $t'=10$ orbits for the MAG\_R1 model.
The gas density and magnetic field exhibit a similar overall appearance to that at $t'=8$ orbits.

\begin{figure}
    \centering
    \includegraphics[width=\linewidth]{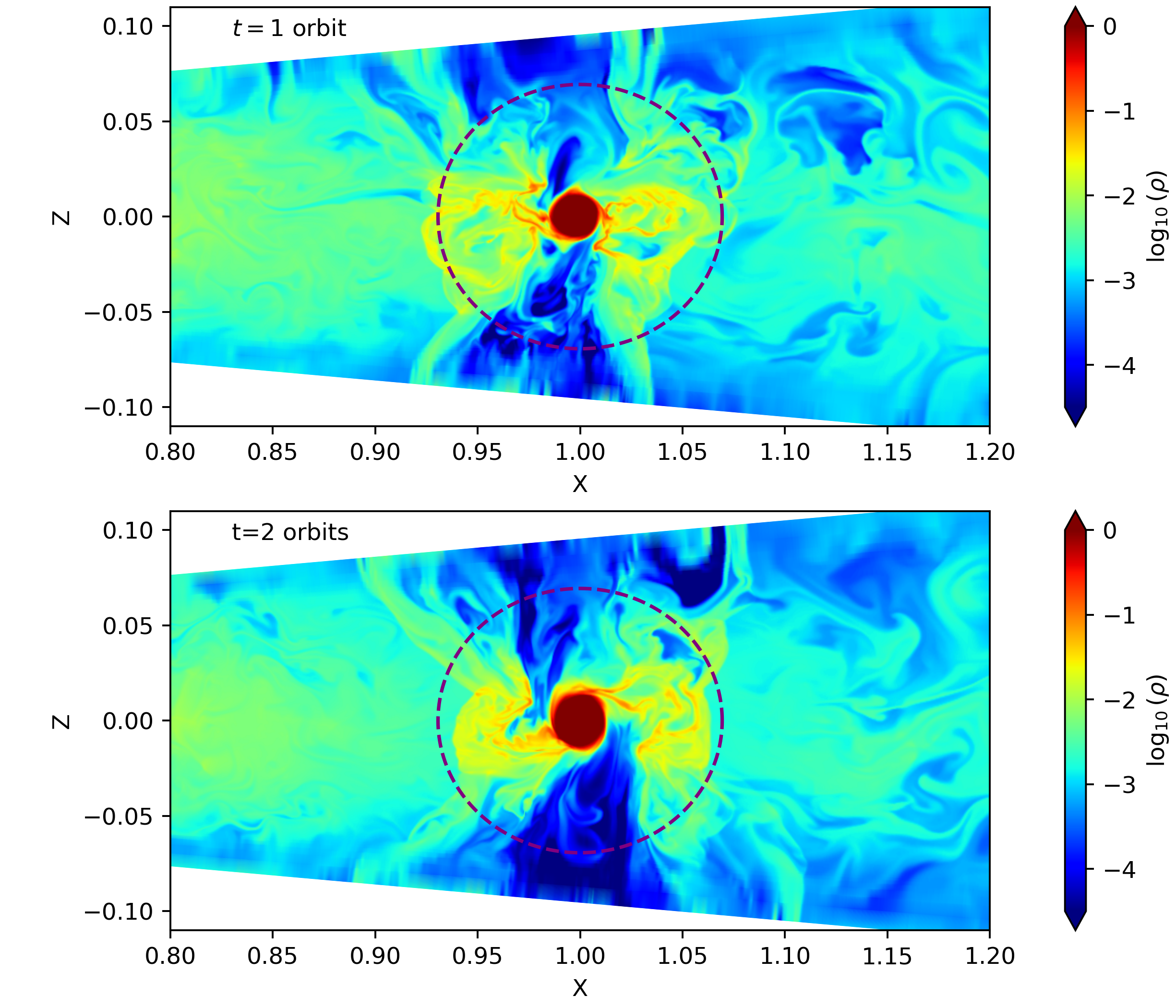}
    \caption{Vertical slices of the CPD density during the first two orbital periods after planet's insertion for MAG\_R1 with $t'=10$ orbits.}
    \label{fig:fields_tempe}
\end{figure}

\begin{figure}
    \centering
    \includegraphics[width=\linewidth]{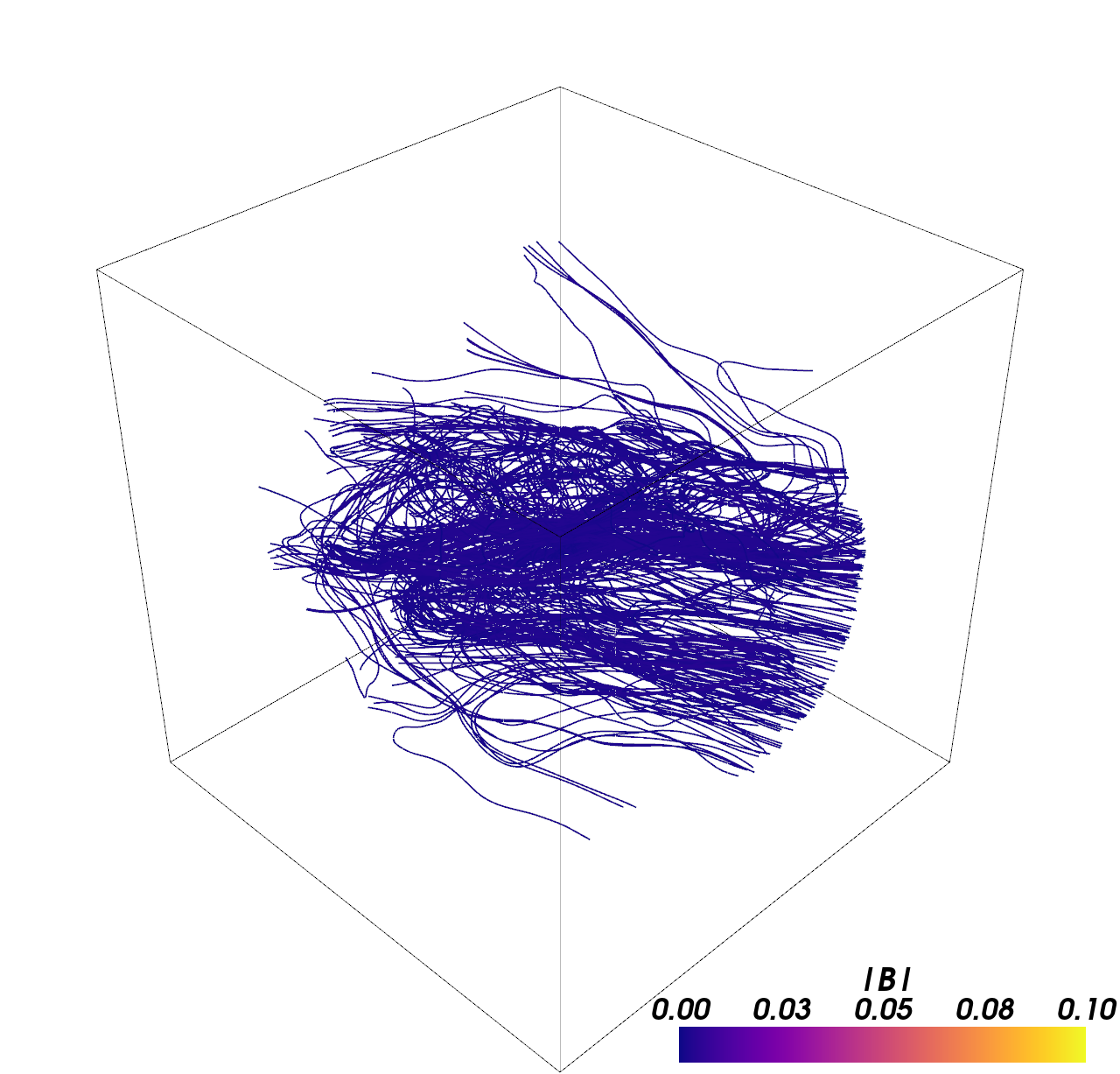}
    \caption{Magnetic field lines in the bipolar-jet formation region in model MAG\_R1 at $t'=10$ orbits.}
    \label{fig:B10}
\end{figure}

Figure \ref{fig:fields_tempe} shows the evolution of the CPD during the first two orbital periods. A comparison of the density maps with those shown in Figure \ref{fig:jets} reveals that the CPD obliquity is similar for planet insertion at $t'=8$ and $t'=10$ orbits. In this case again the cause of the obliquity of the CPD is the development of the toroidal magnetic field $B_{\phi,c}$ (see Fig. \ref{fig:B10} and Section \ref{subsec:tilted}).

Comparison of the maps at $t=2$ orbits shows that the twisting of the bipolar jets occurs more rapidly 
when the planet is inserted at $t'=10$ orbits than at $t'=8$ orbits, despite the MRI having already reached saturation in both cases. Physically, this behavior could be due to the environmental conditions at high latitudes. First, although the statistical properties of the MRI saturate around $t' \approx 3$ orbits, the underlying turbulent dynamo requires additional time to structure the magnetic field. When the planet is inserted into this better-organized magnetic topology at $t'=10$ orbits, the rotational shear of the newly formed CPD can immediately and efficiently wind up these continuous large-scale poloidal lines, accelerating the generation of the helical toroidal component ($B_\phi$).

Second, the sustained turbulent activity between orbits 8 and 10 continuously evacuates gas from the high-latitude, polar regions of the disk. Consequently, the local gas density $\rho$ in the jet launching zone is lower at $t'=10$ orbits. This localized depletion significantly boosts the poloidal Alfvén velocity ($v_{A,p} = B_p/\sqrt{\mu_0 \rho}$), allowing torsional Alfvén waves and magnetic stresses to propagate outward with much less inertial resistance. Therefore, the pre-conditioned environment at $t'=10$ orbits provides the optimal combination of large-scale magnetic "fuel" and lower inertia, resulting in a significantly faster twisting of the bipolar jets post planet insertion.

\end{document}